\documentclass[longauth]{aa} 

\usepackage{graphicx}   
\usepackage{amsmath}    
\usepackage{amssymb}  
\usepackage{multirow}
\usepackage{subfigure}
\usepackage{adjustbox}
\usepackage{xcolor}
\usepackage{float}
\usepackage{amsmath}    
\usepackage{lscape} 
\usepackage{caption}
\usepackage[switch]{lineno}
\usepackage[separate-uncertainty=true, separate-uncertainty-units = single, range-phrase=-, range-units=single]{siunitx}
\usepackage{txfonts}
\usepackage{natbib}
\usepackage[colorlinks=true,linkcolor=blue,allcolors=blue]{hyperref}
\bibpunct{(}{)}{;}{a}{}{,}             

\newcommand{\orcit}[1]{\protect\href{https://orcid.org/#1}{\protect\includegraphics[width=8pt]{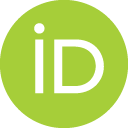}}}

\begin{document}

   \title{Multi-wavelength picture of the misaligned BL Lac object 3C 371}
    \titlerunning{Multi-wavelength picture of 3C 371}

   \author{J. Otero-Santos
          \orcit{0000-0002-4241-5875}\inst{\ref{csic} \and \ref{infn}}\fnmsep\thanks{Contact: \email{joteros@iaa.es}}
          \and
          C. M. Raiteri\orcit{0000-0003-1784-2784}\inst{\ref{oato}}
          \and
          A. Tramacere\orcit{0000-0002-8186-3793}\inst{\ref{univ_geneva}}
          \and
          J. Escudero Pedrosa\orcit{0000-0002-4131-655X}\inst{\ref{csic}}
          \and
          J. A. Acosta-Pulido\orcit{0000-0002-0433-9656}\inst{\ref{iac} \and \ref{ull}}
          \and
          M. I. Carnerero\orcit{0000-0001-5843-5515}\inst{\ref{oato}}
          \and
          M. Villata\orcit{0000-0003-1743-6946}\inst{\ref{oato}}
          \and
          I. Agudo\orcit{0000-0002-3777-6182}\inst{\ref{csic}}
          \and
          I. A. Rahimov\orcit{0000-0002-9185-6239}\inst{\ref{svetloe}}
          \and
          T. S. Andreeva\orcit{0000-0003-3613-6252}\inst{\ref{svetloe}}
          \and
          D. V. Ivanov\orcit{0000-0002-9270-5926}\inst{\ref{svetloe}}
          \and
          N. Marchili\orcit{0000-0002-5523-7588}\inst{\ref{inaf_radio}}
          \and
          S. Righini\orcit{0000-0001-7332-5138}\inst{\ref{inaf_radio}}
          \and
          M. Giroletti\orcit{0000-0002-8657-8852}\inst{\ref{inaf_radio}}
          \and
          M. A. Gurwell\orcit{0000-0003-0685-3621}\inst{\ref{harvard}}
          \and
          S. S. Savchenko\orcit{0000-0003-4147-3851}\inst{\ref{stp} \and \ref{sao} \and \ref{pulkovo}}
          \and
          D. Carosati\orcit{0000-0001-5252-1068}\inst{\ref{ept} \and \ref{inaf-tng}}
          \and
          W. P. Chen\orcit{0000-0003-0262-272X}\inst{\ref{taiwan}}
          \and
          S. O. Kurtanidze\orcit{0000-0002-0319-5873}\inst{\ref{abastumani}}
          \and
          M. D. Joner\orcit{0000-0003-0634-8449}\inst{\ref{wmo}}
          \and
          E. Semkov\orcit{0000-0002-1839-3936}\inst{\ref{sofia_obs}}
          \and
          T. Pursimo\inst{\ref{not} \and \ref{aarhus}}
          \and
          E. Benítez\orcit{0000-0003-1018-2613}\inst{\ref{unam}}
          \and
          G. Damljanovic\orcit{0000-0002-6710-6868}\inst{\ref{vidojevica}}
          \and 
          G. Andreuzzi\orcit{0000-0001-5125-6397}\inst{\ref{inaf-tng}}
          \and
          G. Apolonio\inst{\ref{wmo}}
          \and
          G. A. Borman\orcit{0000-0002-7262-6710}\inst{\ref{CrAO}}
          \and
          V. Bozhilov\orcit{0000-0002-3117-7197}\inst{\ref{sofia_univ}}
          \and
          F. J. Galindo-Guil\orcit{0000-0003-4776-9098}\inst{\ref{cefca}}
          \and
          T. S. Grishina\orcit{0000-0002-3953-6676}\inst{\ref{stp}}
          \and
          V. A. Hagen-Thorn\orcit{0000-0002-6431-8590}\inst{\ref{stp}}
          \and
          D. Hiriart\orcit{0000-0002-4711-7658}\inst{\ref{unam2}}
          \and
          H. Y. Hsiao\inst{\ref{taiwan}}
          \and
          S. Ibryamov\orcit{0000-0002-4618-1201}\inst{\ref{shumen}}
          \and
          R. Z. Ivanidze\orcit{0009-0005-7297-8985}\inst{\ref{abastumani}}
          \and
          G. N. Kimeridze\orcit{0000-0002-5684-2114}\inst{\ref{abastumani}}
          \and
          E. N. Kopatskaya\orcit{0000-0001-9518-337X}\inst{\ref{stp}}
          \and
          O. M. Kurtanidze\orcit{0000-0001-5385-0576}\inst{\ref{abastumani} \and \ref{engelhardt}}
          \and
          V. M. Larionov\inst{\ref{stp}}
          \and
          E. G. Larionova\orcit{0000-0002-2471-6500}\inst{\ref{stp}}
          \and
          L. V. Larionova\orcit{0000-0002-0274-1481}\inst{\ref{stp}}
          \and
          M. Minev\orcit{0000-0002-5702-5095}\inst{\ref{sofia_univ} \and \ref{sofia_obs}}
          \and
          D. A. Morozova\orcit{0000-0002-9407-7804}\inst{\ref{stp}}
          \and
          M. G. Nikolashvili\orcit{0000-0003-0408-7177}\inst{\ref{abastumani}}
          \and
          E. Ovcharov\orcit{0000-0002-0589-0810}\inst{\ref{sofia_univ}}
          \and
          L. A. Sigua\orcit{0000-0002-6410-1084}\inst{\ref{abastumani}}
          \and
          M. Stojanovic\orcit{0000-0002-4105-7113}\inst{\ref{vidojevica}}
          \and
          I. S. Troitskiy\orcit{0000-0002-4218-0148}\inst{\ref{stp}}
          \and
          Yu. V. Troitskaya\orcit{0000-0002-9907-9876}\inst{\ref{stp}}
          \and
          A. Tsai\inst{\ref{taiwan} \and \ref{sunyat-se} \and \ref{tara}}
          \and
          A. Valcheva\orcit{0000-0002-2827-4105}\inst{\ref{sofia_univ}}
          \and
          A. A. Vasilyev\orcit{0000-0002-8293-0214}\inst{\ref{stp}}
          \and
          O. Vince\orcit{0009-0008-5761-3701}\inst{\ref{vidojevica}}
          \and
          E. Zaharieva\orcit{0000-0001-7663-4489}\inst{\ref{sofia_univ}}
          \and
          A. V. Zhovtan\inst{\ref{CrAO}}
          }

   \institute{Instituto de Astrof\'isica de Andalucía (CSIC), Glorieta de la Astronomía s/n, 18008 Granada, Spain \label{csic}
            \and
             Istituto Nazionale di Fisica Nucleare, Sezione di Padova, 35131 Padova, Italy
             \label{infn}
             \and
             INAF-Osservatorio Astrofisico di Torino, via Osservatorio 20, 10025 Pino Torinese, Italy \label{oato}
            \and  
             Department of Astronomy, University of Geneva, Chemin Pegasi 51, 1290 Versoix, Switzerland \label{univ_geneva}
            \and
             Instituto de Astrof\'isica de Canarias (IAC), E-38200 La Laguna, Tenerife, Spain \label{iac}
            \and
             Universidad de La Laguna (ULL), Departamento de Astrof\'isica, E-38206 La Laguna, Tenerife, Spain \label{ull}
            \and
             Institute of Applied Astronomy, Russian Academy of Sciences, St. Petersburg, Russia
             \label{svetloe}
             \and
             INAF – Istituto di Radioastronomia, Via Gobetti 101, 40129 Bologna, Italy \label{inaf_radio}
            \and
             Center for Astrophysics | Harvard \& Smithsonian, 60 Garden Street, Cambridge MA 02138 USA \label{harvard}
             \and
             Saint Petersburg State University, 7/9 Universitetskaya nab., St. Petersburg, 199034 Russia \label{stp}
             \and 
             Special Astrophysical Observatory, Russian Academy of Sciences, 369167, Nizhnii Arkhyz, Russia  \label{sao}
             \and
             Pulkovo Observatory, St. Petersburg, 196140, Russia \label{pulkovo}
             \and
             EPT Observatories, Tijarafe, La Palma, Spain \label{ept}
             \and
             INAF, TNG Fundaci\'on Galileo Galilei, La Palma, Spain \label{inaf-tng}
             \and
             Institute of Astronomy, National Central University, Taoyuan 32001, Taiwan \label{taiwan}
             \and
             Abastumani Observatory, Mt. Kanobili, 0301 Abastumani, Georgia \label{abastumani}
             \and
             Department of Physics and Astronomy, N283 ESC, Brigham Young University, Provo, UT 84602, USA \label{wmo}
             \and
             Institute of Astronomy and National Astronomical Observatory, Bulgarian Academy of Sciences, 72 Tsarigradsko shosse Blvd., 1784 Sofia, Bulgaria \label{sofia_obs}
             \and
             Nordic Optical Telescope, Apartado 474, E-38700 Santa Cruz de La Palma, Santa Cruz de Tenerife, Spain \label{not}
             \and 
             Department of Physics and Astronomy, Aarhus University, Munkegade 120, DK-8000 Aarhus C, Denmark \label{aarhus}
             \and
             Universidad Nacional Aut\'onoma de M\'exico, Instituto de Astronom\'ia, AP 70-264, CDMX 04510, Mexico \label{unam}
             \and
             Astronomical Observatory, Volgina 7, 11060 Belgrade, Serbia \label{vidojevica} 
             \and
             Crimean Astrophysical Observatory RAS, P/O Nauchny, 298409, Russia \label{CrAO}
             \and 
             Department of Astronomy, Faculty of Physics, Sofia University ``St. Kliment Ohridski'', 5 James Bourchier Blvd., BG-1164 Sofia, Bulgaria \label{sofia_univ}
             \and
             Centro de Estudios de Física del Cosmos de Aragón (CEFCA), Plaza San Juan 1, 44001 Teruel, Spain \label{cefca}
             \and
             Universidad Nacional Aut\'onoma de M\'exico, Instituto de Astronom\'ia, AP 106, Ensenada 22800, Baja California, Mexico \label{unam2}
             \and
             Department of Physics and Astronomy, Faculty of Natural Sciences, University of Shumen, 115, Universitetska Str., 9712 Shumen, Bulgaria \label{shumen}
             \and 
             Engelhardt Astronomical Observatory, Kazan Federal University, Tatarstan, Russia \label{engelhardt}
             \and 
             National Sun Yat-sen University, No. 70 Lien-hai Road, Kaohsiung, Taiwan 804201 \label{sunyat-se}
             \and 
             Taiwan Astronomical Research Alliance, Taoyuan 32001, Taiwan \label{tara}
             }

   \date{Received July 8, 2024}

 
  \abstract
   {The BL Lac object 3C 371 is one of the targets regularly monitored by the Whole-Earth Blazar Telescope (WEBT), a collaboration of observers studying blazar variability on both short and long timescales.}
   {We aim to evaluate the long-term multi-wavelength (MWL) behaviour of 3C~371, comparing it with results derived from its optical emission in our previous study. For this, we make use of the multi-band campaigns organised by the WEBT collaboration in optical and radio between January 2018 and December 2020, and of public data from \textit{Swift} and \textit{Fermi} satellites and the MOJAVE Very Large Interferometry programme.}
   {We evaluated the variability shown by the source in each band by quantifying the amplitude variability parameter, and also looked for a possible inter-band correlation using the z-discrete correlation function. We also present a deep analysis of the optical-UV, X-ray, and $\gamma$-ray spectral variability. With the MOJAVE data, we performed a kinematics analysis, looking for components propagating along the jet and calculating its kinematics parameters. We then used this set of parameters to interpret the source MWL behaviour, modelling its broadband spectral energy distribution (SED) with theoretical blazar emission scenarios.}
   {The MWL variability of the source in the UV, X-ray, and $\gamma$-ray bands is comparable to that in optical, especially considering the lower coverage of the first two wavebands. On the other hand, the radio bands show variability of much lower magnitude. Moreover, this MWL emission shows a high degree of correlation, which is compatible with zero lag, again with the exception of the radio emission. The radio VLBI images reveal super-luminal motion of one of the identified components, which we used to set constraints on the jet kinematics and parameters, and to estimate a viewing angle of $\theta=(9.6\pm1.6) \degr$, a Doppler factor of $\delta=6.0\pm1.1,$ and a Lorentz factor of $\Gamma=6.0\pm1.8$. The polarised radio emission was found to be anti-correlated with the total flux, and to follow the same behaviour as the polarised optical radiation. The optical-UV spectral behaviour shows a mild harder-when-brighter trend on long timescales, and other trends such as redder-when-brighter on shorter timescales. We
successfully modelled the broadband emission  with a leptonic scenario, where we compared the low and high emission states during the period of complete MWL coverage. The difference between these two states can be ascribed mainly to a hardening of the distribution of particles. The derived features of the source confirm that 3C~371 is a BL~Lac whose jet is not well aligned with the line of sight.}
   {}


   \maketitle
%

\section{Introduction}
Blazars, which are active galactic nuclei (AGN) with one of their jets pointing towards our line of sight, exhibit profuse emission over the entire electromagnetic spectrum. Depending on whether or not they show an optical spectrum that contains features with $|EW|>5$~\AA~in the rest frame \citep{stickel1991}, they are classified into BL Lacertae (BL Lac) objects or flat-spectrum radio quasars \citep[FSRQs;][]{urry1995}, respectively. Their spectral energy distribution (SED) is known for its double-bump shape \citep{abdo2010}, owing to the two physical mechanisms typically invoked to explain their emission. There is consensus over the theory that the low-energy bump is produced by synchrotron radiation \citep[see e.g.][]{koenigl1981}. The high-energy bump is most commonly interpreted within the leptonic inverse Compton (IC) scenario, either via synchrotron self-Compton \citep[SSC; e.g.][]{maraschi1992} or external Compton \citep[EC, see][]{dermer1993}  scattering. However, hadronic processes have gained importance, especially given results suggesting that blazars could be sources of the extragalactic high-energy neutrinos detected by the IceCube Observatory \citep{ice2018a,ice2018b}.

The main feature of blazar multi-wavelength (MWL) emission is its remarkable and unpredictable variability in all wavelengths, and on all possible timescales \citep{wagner1995}. Variability on timescales of larger than a few months or years is usually referred to as long-term variability (LTV). Moreover, timescales of the order of several days to weeks are considered as short-term variability (STV). Finally, blazars also display very fast flux variations that take place over timescales of a few hours or even a few minutes, which is known as intraday variability (IDV). These timescales, present in the broadband emission of blazars and thought to be caused by different physical mechanisms owing to their different nature, are typically entangled, which prevents the straightforward interpretation of this observed variability. Therefore, characterising and disentangling the variability signatures of blazars provides a means to understand and interpret the physical origin of the emission observed from these objects.

Here we aim to study the broadband emission of 3C~371, a blazar belonging to the BL Lac object subclass located at a redshift of $z=0.0510 \pm 0.0003$ \citep{degrijp1992}. As concluded by \cite{wrobel1990}, 3C~371 is a BL Lac object observed under a sufficiently large viewing angle that the two radio lobes are visible and do not overlap. Recently, this source has shown remarkable periods of fast IDV in the optical band, which were revealed thanks to observations performed by the TESS satellite over almost 1 year, with a two-minute observing cadence. In parallel, the Whole-Earth Blazar Telescope (WEBT) collaboration monitored the source extensively in the
optical and radio between 2018 and 2021. These optical data were analysed in {\citet[][hereafter Paper I]{otero-santos2024}}, where we characterised the IDV and LTV signatures in the emission of 3C~371. Complementing these optical data, several MWL instruments and telescopes observed 3C~371 in this same time window, allowing a broadband interpretation of the emission from radio to $\gamma$ rays.

In this work, we characterise the broadband variability and SED of 3C~371 using data over the entire spectrum. The paper is structured as follows: In Sect.~\ref{sec2} we describe the MWL datasets and instruments used in the analysis. In Sect.~\ref{sec3} we present the MWL light curves, evaluating the variability and correlations. We present our analysis of the jet features and jet kinematics in Sect.~\ref{sec4}, and a spectral analysis of optical-UV, X-ray, and $\gamma$-ray bands in Sect.~\ref{sec5}. In Sect.~\ref{sec6}, we analyse the variability and behaviour of the radio and optical polarised emission, and in Sect.~\ref{sec7} we model the broadband SED and describe our interpretation of the results. Some final remarks and conclusions are provided in Sect.~\ref{sec8}.

\section{Observations and data reduction}\label{sec2}

\subsection{Gamma-ray observations: \textit{Fermi}-LAT}\label{sec2.3}
We analysed data from the pair-conversion Large Area Telescope (LAT) on board the \textit{Fermi} satellite. The LAT monitors the $\gamma$-ray sky every three hours in the energy range of 20~MeV to $\sim$2~TeV, and exhibits maximum sensitivity below a few GeV \citep{atwood2009}. For the present data analysis, a region of interest (ROI) with a radius of 15$^{\circ}$ was selected, centred at the position of 3C~371. The data between 2018 and 2021 were retrieved from the LAT Data Server\footnote{\url{https://fermi.gsfc.nasa.gov/ssc/data/access/lat/}}. These data were analysed by selecting all the Pass8 \texttt{P8R3 source} events between 100~MeV and 300~GeV with version 1.2.23 of the standard software \textsc{Fermitools}. A zenith angle cut of 90$^{\circ}$ was introduced to reduce the contamination from the limb of the Earth. Moreover, we used the recommended Galactic diffuse emission model and the isotropic component for the event selection performed here\footnote{\url{https://fermi.gsfc.nasa.gov/ssc/data/access/lat/BackgroundModels.html}}.

First, we analysed one year of data with a binned likelihood method to obtain the model that describes the sources in the field. For this analysis, we included not only all the sources within the ROI but also the sources contained in an extra annular region of 10$^{\circ}$ in radius. The spectral shape and normalisation of the sources within the ROI are left as free parameters in the model, while they were fixed to the catalogue values for those in the annular region. For this, we used the 4FGL-DR2 catalogue \citep{abdollahi2020,ballet2020}. The normalisation of the diffuse components was also left free in the iteration used to obtain the model file. In order to converge to the final model, all the sources with a test statistics (TS) of <2 were discarded. 

\begin{figure}
        \centering
        \includegraphics[width=0.93\columnwidth]{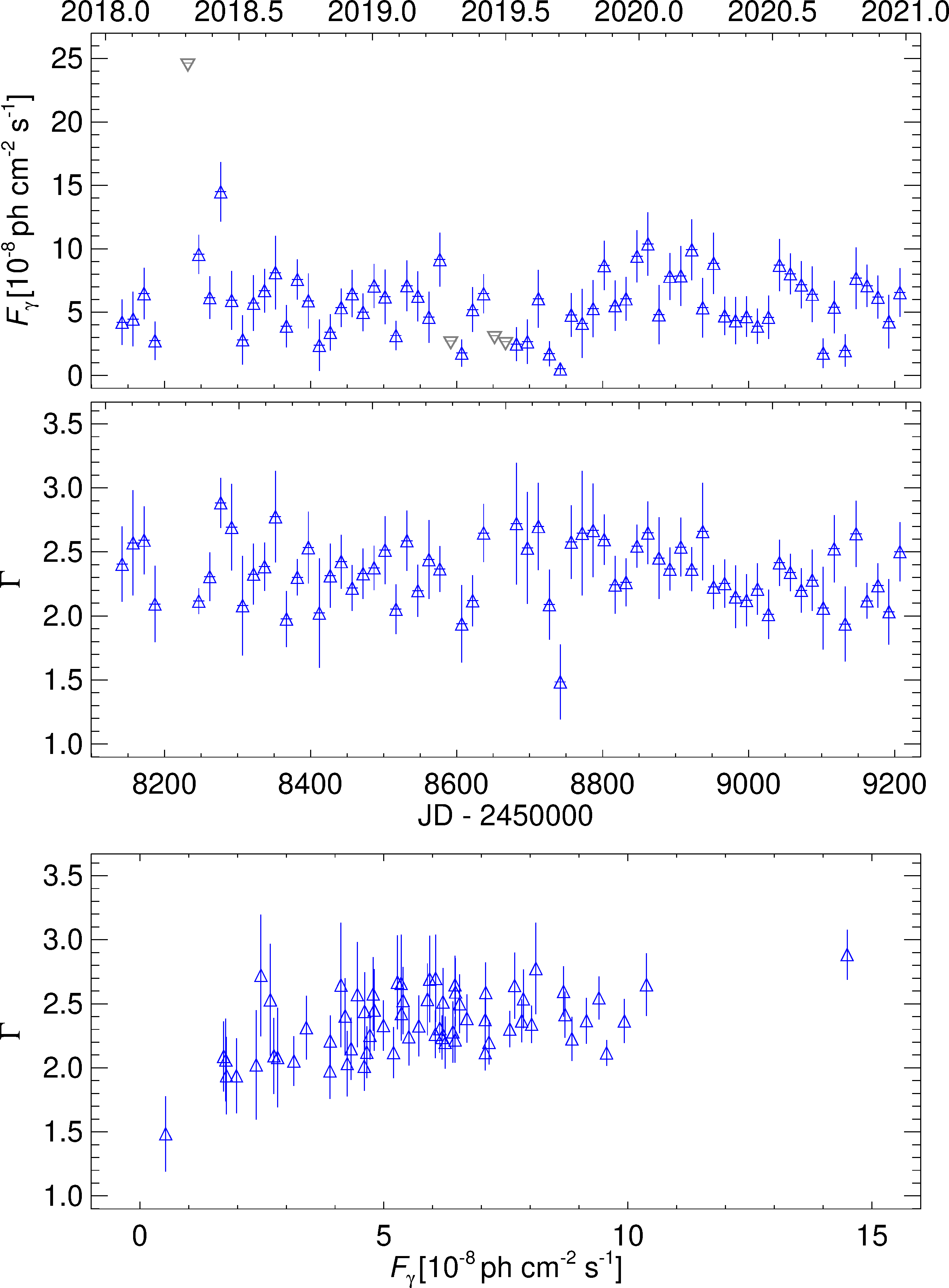}
    \caption{{\textit{Fermi}-LAT $\gamma$-ray light curve of 3C 371.} \textit{Top}: 15 day binned $\gamma$-ray flux light curve between 100~MeV and 300~GeV obtained from the \textit{Fermi}-LAT data. \textit{Middle}: $\gamma$-ray spectral index as a function of time. \textit{Bottom}: $\gamma$-ray spectral index as a function of $\gamma$-ray flux.}
    \label{fig:fermi_lcs}
\end{figure}

To obtain the light curve, we used the resulting model as input for an unbinned likelihood analysis on each 15 day data bin. This bin width was selected due to the relatively faint emission of 3C~371 in the $\gamma$-ray band, which led to mostly upper limits on shorter timescales. We assumed a power-law spectral shape for each bin, leaving both the spectral index and normalisation as free parameters. Finally, the parameters of all the point-like sources were fixed, except for five sources located at a distance of <10$^{\circ}$ from 3C~371 with TS>25 and a variability index as reported in the 4FGL-DR2 catalogue of higher than that of 3C~371. The $\gamma$-ray flux light curve and the spectral index evolution with time are shown in Fig.~\ref{fig:fermi_lcs}.

\subsection{X-ray observations: \textit{Swift}-XRT}
\begin{figure}
        \centering
        \includegraphics[width=0.93\columnwidth]{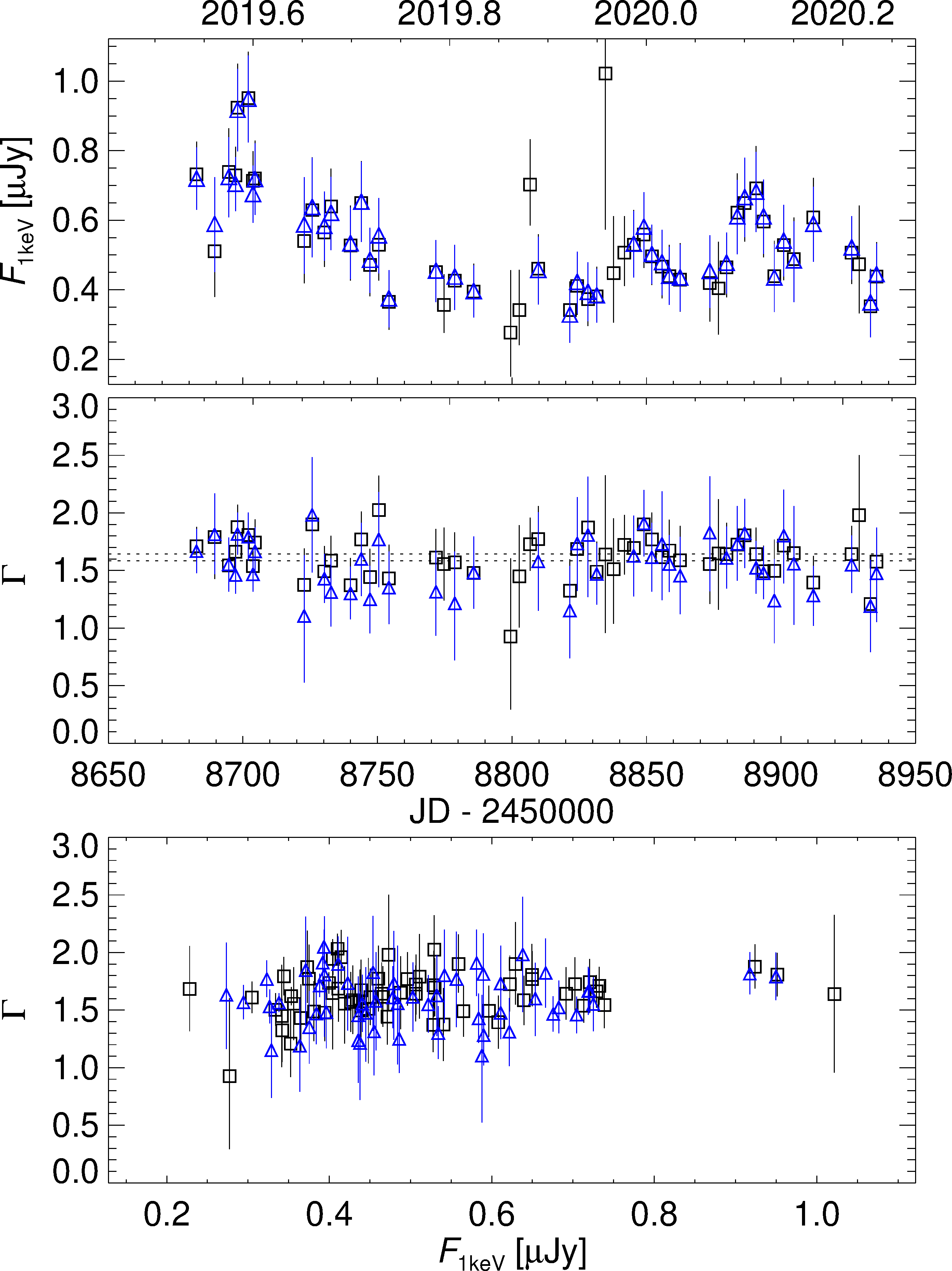}
    \caption{{\textit{Swift}-XRT X-ray light curve of 3C 371.} \textit{Top}: X-ray flux densities at 1 keV from \textit{Swift}-XRT obtained with both the Cash (black squares) and the chi-squared (blue triangles) statistics. \textit{Middle}: Power-law index $\Gamma$ as a function of time. \textit{Bottom}: Power-law index $\Gamma$ as a function of 1 keV flux density.}
    \label{fig:xrt_lcs}
\end{figure}

The \textit{Swift} satellite observed 3C~371 in X-rays with its X-Ray Telescope \citep[XRT; see][]{burrows2004} between approximately MJD~58682 and MJD~58935. 
\textit{Swift} data were reduced with the {\texttt HEASoft} package\footnote{\url{https://heasarc.gsfc.nasa.gov/lheasoft/}} version 6.30.1 with calibration files updated as of November 2022. Observations in pointing mode were processed with the  {\texttt xrtpipeline} task, selecting grades 0--12 for the photon counting (PC) mode
and grades 0--2 for the windowed timing (WT) mode. 
In the period of interest, there were 54 PC observations and only two short WT exposures, which we do not consider hereafter. The count rates were low, and so pile-up was not an issue. We extracted the source counts in a circular region with 50 arcsec aperture centred on the source; the background was estimated from a close-by source-free circular region with a radius of 80 arcsec. We used the package {\texttt Xspec} to fit the spectra in the 0.3--10 keV energy range with an absorbed power law. The hydrogen column density was fixed to the Galactic value of $N_{\rm H}=4.11 \times 10^{20} \rm \, cm^{-2}$ \citep{kalberla2005} with elemental abundances from \citet{wilms2000}. 

Because of the low number of counts, {in order to have a stability cross-check of the analysis,} we used both the Cash statistics and the chi-squared statistics, and in this case we binned the spectra to have at least 20 counts per energy bin. {For all the analyses presented in this paper, we used the results of the Cash statistics, because this method is more sensitive than the chi-squared analysis in situations where observations  are affected by low number counts, and can recover more flux points \citep{cash1979}.}
Figure~\ref{fig:xrt_lcs} shows the XRT flux density light curve at 1 keV obtained with both statistics, together with the power-law index $\Gamma$, which shows no trend with brightness.


\subsection{Optical-UV observations: \textit{Swift}-UVOT}
\begin{figure}
    \centering
    \includegraphics[width=0.93\columnwidth]{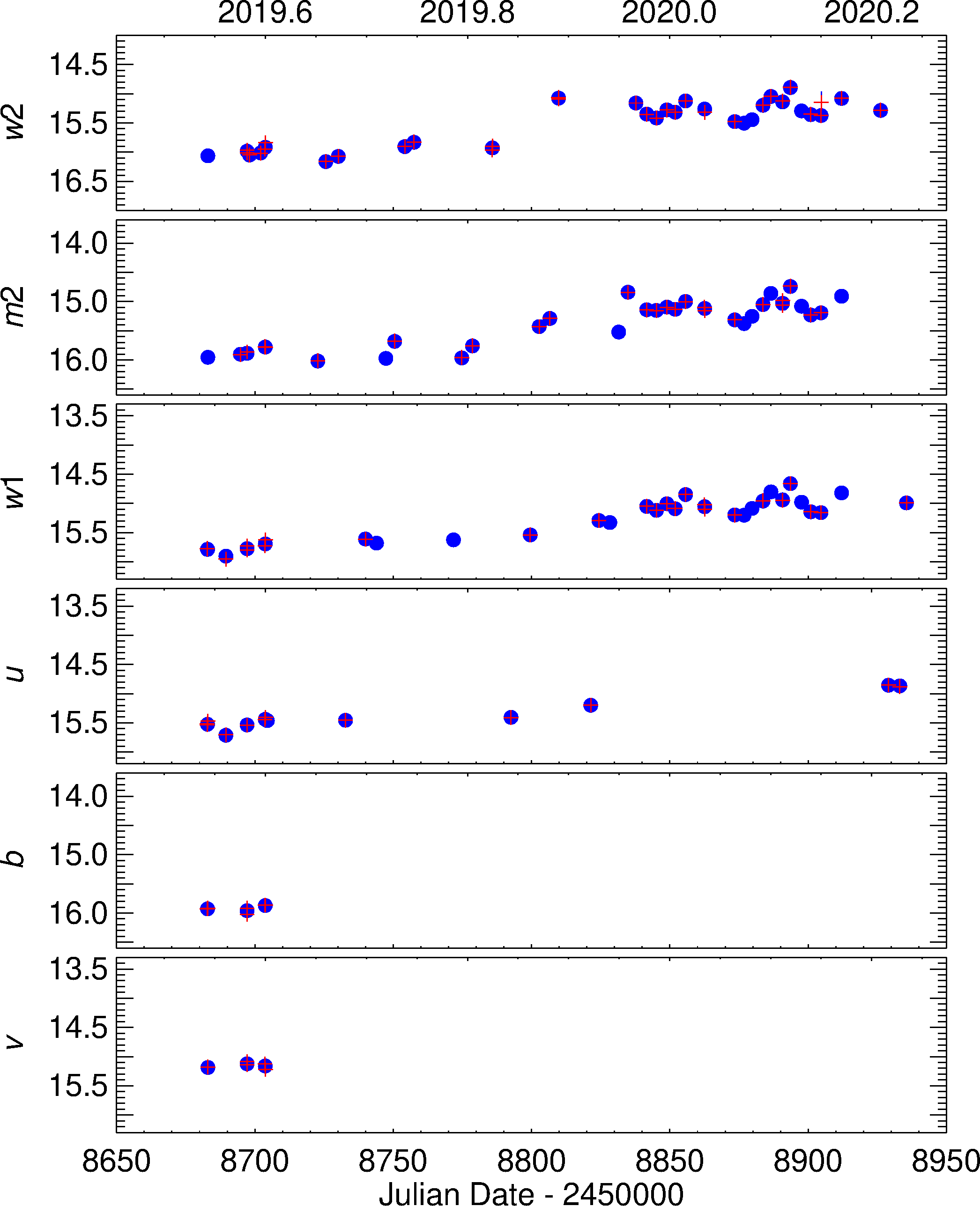}
    \caption{{\it Swift}-UVOT optical--UV light curves (observed magnitudes) in the period considered in this paper. Red plus symbols correspond to photometry on single exposures; blue dots correspond to the results of co-adding the exposures of the same observations.}
    \label{fig:uvot_lcs}
\end{figure}

The Ultra-Violet and Optical Telescope \citep[UVOT, see][]{roming2005} on board the \textit{Swift} satellite observed 3C~371 during the same period as XRT with three optical filters ($v$, $b,$ and $u$)
and three UV filters 
($w1$, $m2,$ and $w2$).
A total number of 3, 3, and 10 observations are available for the $v$, $b,$ and $u$ filters, respectively, in the period considered here. On the other hand, the UV filters have a better coverage, with 28, 30, and 30 observations in the $w1$, $m2,$ and $w2$ filters, respectively. 
The UVOT data were processed using a 5$\arcsec$ aperture radius for the source extraction and an annulus with radii of  15$\arcsec$ and 25$\arcsec$  centred on the source for the background. We derived the source photometry both for each single exposure of the same observation and on the co-added exposures. In the former case, we used the {\texttt uvotmaghist} task, while in the latter case we first checked the image alignment before co-adding the exposures with the task {\texttt uvotimsum,} and finally performed the photometry with the task {\texttt uvotsource}. The resulting light curves in both cases are shown in Fig.~\ref{fig:uvot_lcs}.


We also evaluated possible calibration effects in the counts-to-flux conversion for each filter. To this end, we calculated the $b-v$ index for all observations, finding that they deviate from the validity range provided by \cite{breeveld2010}, which was originally calculated for Pickles stars and gamma-ray bursts (GRBs). Therefore, the counts-to-flux conversion factors 
are not suitable for the spectral type of our source. In order to recalibrate the effective wavelengths and conversion factors for each filter, we followed the prescriptions by \cite{raiteri2010}. We constructed an averaged optical-UV SED, which was later convoluted with the effective areas of the UVOT filters to obtain the new effective wavelengths and counts-to-flux conversion factors. The new effective wavelengths were found to be 5431~\AA, 4362~\AA, 3478~\AA, 2627~\AA, 2258~\AA, \ and 2092~\AA \ for the $v$, $b$, $u$, $w1$, $m2,$ and $w2$ filters, respectively. Moreover, the counts-to-flux conversion rates for these filters were estimated to be 2.602, 1.467, 1.647, 4.392, 8.356, and 5.968, in units of $10^{-16}$, respectively.

In order to correct the observations performed by UVOT from Galactic extinction, we calculated the interstellar extinction in the effective wavelengths reported above using the interstellar extinction curve from \cite{cardelli1989} and the procedure from \cite{raiteri2010}. We obtained the Galactic extinction values of $A_{v}=0.095$, $A_{b}=0.124$, $A_{u}=0.150$, $A_{w1}=0.210$, $A_{m2}=0.261$, and $A_{w2}=0.248$ for the different bands. 

Finally, we corrected the data for the contribution of the host galaxy. To estimate its value in each UVOT filter, we adopted the 13 Gyr elliptical galaxy template by \cite{polletta2007} and calculated the ratio of the galaxy flux in the UVOT filters with respect to that in the $BVRI$ fluxes.
Considering the aperture of 5$\arcsec$ used for the UVOT data analysis, we estimate that $\sim$35.6\% {of the total host galaxy emission is contributing to the observed flux}. The contributions of the host galaxy to the different optical-UV bands in flux density units are reported in Table~\ref{tab:host_galaxy_contributions}. These contributions were subtracted from the observed flux density values.

\begin{table}
\begin{center}
\caption{Contribution of the host galaxy to the total observed flux in the different \textit{Swift} optical and UV bands.}
\label{tab:host_galaxy_contributions}
\resizebox{\columnwidth}{!}{%
\begin{tabular}{ccccc}
\hline
\multirow{2}{*}{Band} &  \multirow{2}{*}{$\lambda_{eff}$ [\AA]} & \multirow{2}{*}{Aperture [$\arcsec$]} & Containment  & Contained \\
                  &  &  & fraction [\%] & flux$^{a}$ [mJy] \\ \hline
  $v$     &     5431       &    5.0      &          35.6            &     1.467  (1.601)       \\ \hline
  $b$     &     4362       &    5.0      &          35.6            &     0.595  (0.667)       \\ \hline
  $u$     &     3478       &    5.0      &          35.6            &     0.164  (0.188)       \\ \hline
  $w1$     &    2627        &   5.0       &         35.6             &    0.029  (0.036)        \\ \hline
  $m2$     &    2258        &   5.0       &         35.6             &    0.008  (0.011)        \\ \hline
   $w2$    &    2092        &   5.0       &         35.6             &    0.009  (0.011)        \\ \hline
\end{tabular}
}
\end{center}
\vspace{-0.2cm}
Notes: $^a$Values outside (between) parentheses correspond to the host galaxy contribution before (after) correcting for the Galactic extinction reddening effect. 
\end{table}

\subsection{WEBT optical observations}
The optical data presented in Paper I were also included as part of the monitoring performed by the WEBT\footnote{\url{https://www.oato.inaf.it/blazars/webt/}} on 3C~371. These data include photometric data in the Johnson--Cousins $BVRI$ bands and polarimetric data in the $R$ band provided by several observatories and telescopes around the world, as indicated in Table~1 of Paper I. All the data have been corrected for the reddening effect of Galactic extinction {and the contribution of the host galaxy to the total flux in each band was subtracted}. Further details of the analysis {are provided} in Paper I. 

\subsection{Radio observations}
\begin{figure}
        \centering
        \includegraphics[width=0.93\columnwidth]{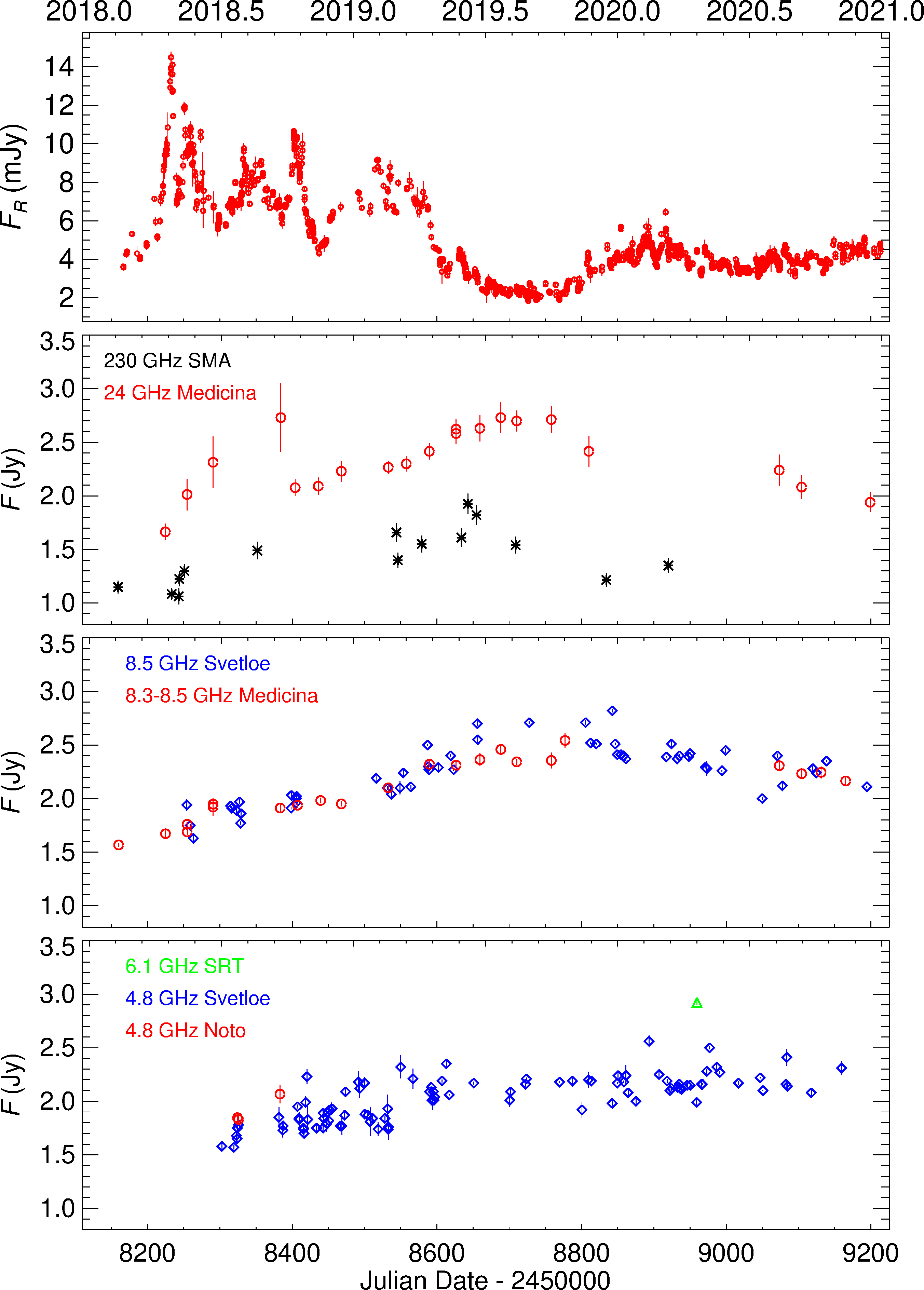}
    \caption{Radio light curves of 3C~371 compared with the dereddened and host-subtracted optical $R$-band flux density (top panel). Different colours and symbols
are used to distinguish the contributing datasets, as specified in Table \ref{tab:WEBT_optical_radio_obs}.}
    \label{fig:radio_LCs}
\end{figure}

Radio observations coordinated as part of the WEBT observing campaigns have also been performed at 4.8~GHz, 6.1~GHz, 8.5~GHz, 24~GHz, and 230~GHz with several radio telescopes, as listed in Table~\ref{tab:WEBT_optical_radio_obs}. The flux density calibrated radio light curves are presented in Fig. \ref{fig:radio_LCs} in comparison with the dereddened and host-subtracted optical $R$-band light curve in flux density units. The data reduction was carried out following a standard procedure \citep{terasranta1998,giroletti2020}. 


\begin{table*}
\begin{center}
\caption{WEBT observatories supporting the observing campaign of 3C~371 in the radio bands.}
\label{tab:WEBT_optical_radio_obs}
\begin{tabular}{llccccc}
\hline
            \multicolumn{7}{c}{\it Radio}\\
Observatory & Country & Telescope size (m) & Band (GHz) & N & Symbol & Colour \\
\hline
Cagliari (SRT) & Italy & 64 & 6.1 & 1 & {$\triangle$} & green\\
Mauna Kea (SMA) & US & $8 \times 6^a$ & 230 & 15 & {\LARGE $\ast$} & black\\
Medicina & Italy & 32 & 8.3, 8.5, 24 & 42 & {\LARGE $\circ$} & red\\
Noto & Italy & 32 & 4.8 & 3 & {\LARGE $\circ$} & red\\
Svetloe & Russia & 32 & 4.8, 8.5 & 155 & {\LARGE $\diamond$} & blue\\
\hline
\end{tabular}\\
\end{center}
\vspace{-0.2cm}
Notes: The telescope size is reported (in cm), together with the total number of observations provided and the symbols and colours used in the light-curve plots. $^a$Radio interferometer composed of eight 6-m dishes.\\

\end{table*}

Visual inspection of Fig.~\ref{fig:radio_LCs} shows that, despite the lower observing cadence of the radio data, the radio emission does not follow the same variability pattern as the optical light curve. While the optical $R$ band goes through a period of fast flares and noticeable variability, the radio emission shows a much slower and less intense variability, with an increasing trend that reaches its maximum when the optical emission is at its minimum. {This cannot be extended to the 230-GHz band, where the lower observing cadence prevents us from drawing any strong conclusion as to the evolution of the flux at this frequency.} We discuss the behaviour and variability of the radio emission further in in Sect.~\ref{sec3}.

\subsection{Radio Very Large Base Array (VLBA) observations}
The Monitoring Of Jets in Active galactic nuclei with VLBA Experiments (MOJAVE)\footnote{\url{https://www.cv.nrao.edu/MOJAVE/index.html}} is a long-term AGN monitoring programme performing Very Large Baseline Interferometry (VLBI) 15~GHz radio observations with the goal of studying the kinematics of extragalactic relativistic jets \citep{lister2018}. The programme has been providing long-term coverage of AGN jets on parsec scales for over 100 sources since 2002.

We present a set of 18  VLBI ultra-high (milliarcsecond) resolution images of the jet in 3C~371 at 15 GHz taken by the MOJAVE collaboration from August 15, 2019, to February 5, 2021. The visibility data, already fully calibrated both in phase and amplitude for the total flux and linear polarisation, were taken from the publicly available MOJAVE archive\footnote{\url{https://www.cv.nrao.edu/MOJAVE/allsources.html}}. {We performed no further calibration of the data.} Then, following the same procedure as in \cite{escudero2024}, we processed the calibrated visibility file from every single observing epoch using \texttt{difmap} \citep[see][]{shepherd1997} to produce the I, Q, and U Stokes’ parameter images. We assumed no circular polarisation emission to be produced by the source. We produced the linear polarisation flux and angle maps from the Q and U images, respectively. After mapping, the total-flux brightness distribution of the jet for all the observing epochs was modelled using \texttt{difmap} in the visibility plane to describe every two-dimensional total-flux image as a set of circular Gaussian emission components. Each of these components is characterised by a position, a full width at half maximum (FWHM), and an associated brightness. By identifying every one of the modelled Gaussian components over time, we were then able to follow the time evolution of the most prominent emission regions in the jet of 3C~371 (see Sect.~\ref{sec4}). {The model fit results from this analysis and the estimates of the radio polarisation are included in Appendices~\ref{sec:appendixA} and \ref{sec:appendixB}, respectively.}

\section{Multi-wavelength light curves}\label{sec3}

\subsection{Variability}
\begin{figure}
        \centering
        \includegraphics[width=0.9\columnwidth]{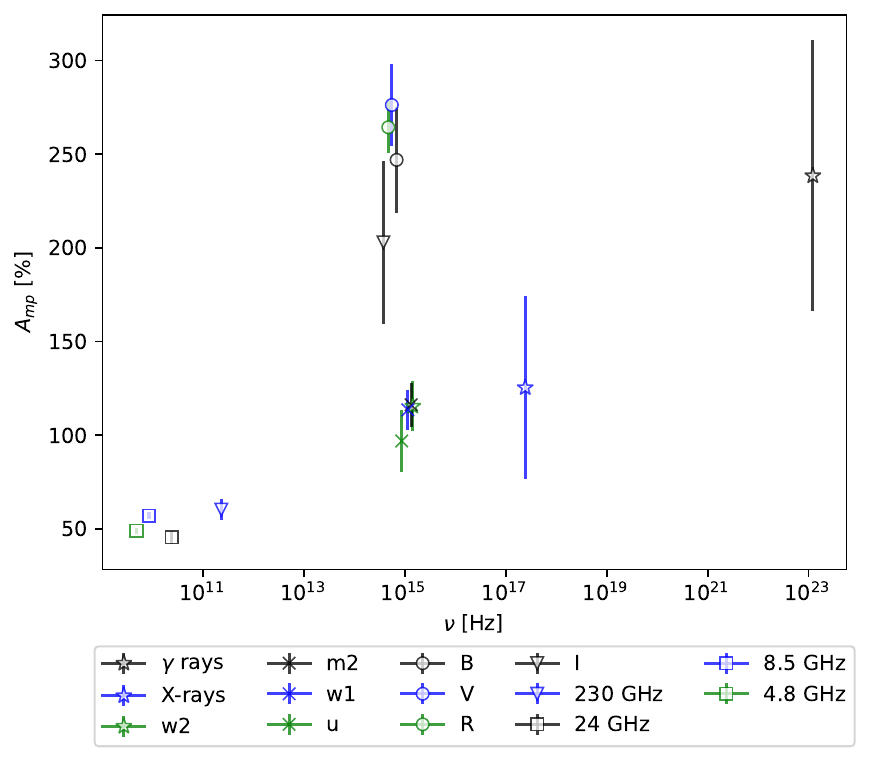}
    \caption{Amplitude variability for the different bands as a function of frequency. Different markers and colours represent different bands of the spectrum, as indicated in the legend.}
    \label{fig:amp}
\end{figure}

We evaluated the LTV of the MWL light curves, comparing them to the variability estimated for the optical bands presented in Paper I. To this end, we used the amplitude variability parameter \citep[see][]{romero1999}, which quantifies the amount of variability shown by each dataset as
\begin{equation}
A_{mp}(\%)=\frac{100}{<m>}\sqrt{(m_{\text{max}}-m_{\text{min}})^{2}-2\sigma^{2}},
\label{eq:var_amp}
\end{equation}
where $<m>$ is the mean flux density value {of the complete dataset}, $m_{\text{max}}$ and $m_{\text{min}}$ are the maximum and minimum measurements, and $\sigma$ is the observational uncertainty. The uncertainty of $A_{mp}$ can be expressed as
\begin{equation}
\begin{split}
\Delta A_{mp}(\%)=100 \times \left(\frac{m_{\text{max}}-m_{\text{min}}}{<m>A_{mp}}\right) \times \\ 
\times \sqrt{\left( \frac{\sigma_{\text{max}}}{<m>} \right)^{2} + \left( \frac{\sigma_{\text{min}}}{<m>} \right)^{2} + \left( \frac{\sigma_{<m>}}{m_{\text{max}}-m_{\text{min}}} \right)^{2}A_{mp}^{4}}.
\end{split}
\label{eq:var_amp_err}
\end{equation}
{In this equation, $\sigma_{<m>}$ represents the standard deviation of all flux measurements.} The results for the different bands are gathered in Table~\ref{tab:ltv} and are represented in Fig.~\ref{fig:amp}. The values obtained for the optical $BVRI$ bands are also included for comparison.

\begin{table}
\begin{center}
\caption{Results of the LTV analysis performed on the MWL emission of 3C~371.}
\label{tab:ltv}

\begin{tabular}{cccccc}
\hline
\multirow{2}{*}{Band} & $F_{min}$ & $F_{max}$ & $\Delta F_{max}$  & $A_{mp}$  & $\Delta A_{mp}$ \\
 & [mJy] & [mJy] & [mJy] & [\%] & [\%] \\ \hline
$\gamma$ rays$^{a}$  & 0.53 & 14.49 & 13.96 & 238.5 & 72.4 \\ \hline
$1$~keV$^{b}$  & 0.28 & 0.95 & 1.20 & 125.4 & 48.8 \\ \hline
$w2^{c}$  & 0.33 & 1.07 & 0.74 & 115.5 & 13.3 \\ \hline
$m2^{c}$  & 0.38 & 1.25 & 0.87 & 116.2 & 11.7 \\ \hline
$w1^{c}$  & 0.47 & 1.55 & 1.08 & 113.5 & 10.7 \\ \hline
$u^{c}$  & 0.67 & 1.70 & 1.03 & 96.9 & 16.7 \\ \hline
$b^{c}$  & 1.24 & 1.40 & 0.16 & -- & -- \\ \hline
$v^{c}$  & 1.77 & 1.97 & 0.20 & -- & -- \\ \hline
$B^{c,d}$  & 0.85 & 6.92 & 6.08 & 247.0 & 28.4 \\ \hline
$V^{c,d}$  & 1.16 & 10.09 & 8.92 & 276.3 & 21.9 \\ \hline
$R^{c,d}$  & 1.84 & 14.49 & 12.65 & 264.4 & 13.9 \\ \hline
$I^{c,d}$  & 2.38 & 16.15 & 13.78 & 203.0 & 43.5 \\ \hline
230~GHz$^{e}$  & 1.06 & 1.93 & 0.87 & 60.3 & 5.7 \\ \hline
24~GHz$^{e}$  & 1.66 & 2.73 & 1.07 & 45.4 & 3.6 \\ \hline
8.5~GHz$^{e}$  & 1.57 & 2.82 & 1.25 & 57.0 & 1.8 \\ \hline
4.8~GHz$^{e}$  & 1.57 & 2.56 & 0.99 & 49.0 & 1.7 \\ \hline
\end{tabular}
\end{center}
{\flushleft
\vspace{-0.2cm}
\hspace{0.5cm}$^{a}$Flux of the $\gamma$-ray band in units of $10^{8}$~cm$^{-2}$~s$^{-1}$.\\
\hspace{0.5cm}$^{b}$Flux of the X-ray 1~keV light curve in units of $\mu$Jy.\\
\hspace{0.5cm}$^{c}$Dereddened and host-galaxy-subtracted flux density values.\\
\hspace{0.5cm}$^{d}$Estimations from Paper I.\\
\hspace{0.5cm}$^{e}$Flux of the radio bands in units of Jy.\\
}
\end{table}

We observe that the difference between the maximum and minimum flux is maximum in the optical and $\gamma$-ray bands, which also show the maximum values of $A_{mp}$. On the other hand, the variability shown by the rest of the bands is clearly lower, with the caveat that the UV and X-ray data have a much shorter time coverage than the other bands. However, and despite this limitation, the variability of the UV and X-ray bands is closer to that of the optical bands than that of the radio data. The $v$ and $b$ UVOT bands have only three measurements during the considered period, which is insufficient to evaluate their variability. The radio bands clearly show variability of   a lower amplitude than that observed for the MWL emission, despite the fact that the data cover approximately the same time span as the optical datasets. This is a common result for $\gamma$-ray blazars, which usually show the least variability in the radio bands \citep[see e.g.][]{ahnen2017,ahnen2018,magic2023}.

\subsection{Correlations}
Understanding how the emissions in the different bands evolve and behave with respect to each other can provide key information on, for instance, the structure of the jet and the location of the emitting region(s) \citep[see e.g.][]{max-moerbeck2014,liodakis2018,liodakis2019}. For this reason, we performed a correlation analysis to evaluate the variation and evolution of the emission in the different bands. 
To this end, we used the z-discrete correlation function (ZDCF) implementation from \cite{alexander2013}. In particular, we employed the \textsc{python} package \texttt{pyZDCF}\footnote{\url{https://github.com/LSST-sersag/pyzdcf}} implemented by \cite{jankov2022}, which is based on the original \textsc{fortran} code developed by \cite{alexander2013}. 
As reference, we took the WEBT $R$-band light curve with the best sampling among the long-term light curves presented in Sect.~\ref{sec2}, performing a cross-correlation function (CCF) using the light curves corresponding to the different bands with respect to the optical $R$ band emission. Owing to the definition of the CCF, positive (negative) time lags mean that the second data series is lagging (leading) the $R$-band light curve. 

In order to assess the statistical significance of the results, we applied the prescription from \cite{max-moerbeck2014b}, which is widely used in correlation studies \citep[see e.g.][]{max-moerbeck2014b,raiteri2023,magic2023}. This method is based on the simulation of artificial light curves with statistical properties that mimic those from the real datasets as closely as possible. In this case, we produced a large number of simulated light curves (10$^{4}$ simulations) using the light-curve simulation method from \cite{emmanoulopoulos2013}, which makes use of the power spectral distribution (PSD) and probability density function (PDF) of the real data to produce the artificial time series, achieving the best solution in terms of statistical resemblance. In particular, we used the \textsc{python} implementation \texttt{DELCgen}\footnote{\url{https://github.com/samconnolly/DELightcurveSimulation}} developed by \cite{connolly2015} following the instructions provided by \cite{emmanoulopoulos2013}. Finally, for the estimation of the uncertainty of the measured time lag, we implemented the approach of \cite{peterson1998,peterson2004}. This formalism is based on an independent Monte Carlo (MC) randomisation and a random subset selection, deriving the time lag of maximum correlation as the centroid $r_{\text{cen}}$ of all the significant correlations (above a 2$\sigma$ threshold) that also fulfill $r > 0.8 R_{\text{max}}$, and its uncertainty as the 68\% confidence limit of the estimation of $r_{\text{cen}}$.

\begin{figure*}
\centering
        \includegraphics[width=\textwidth]{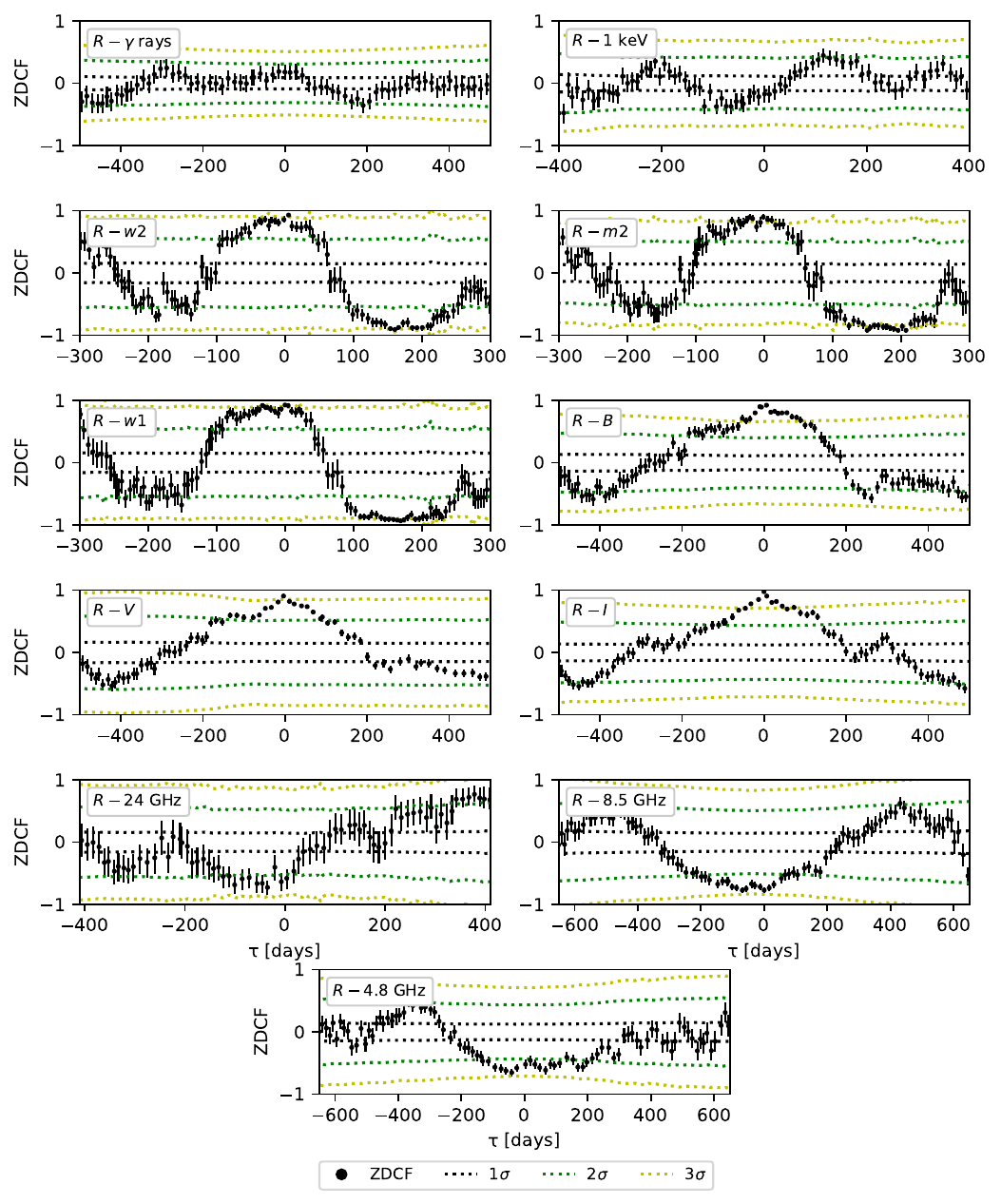}
    \caption{ZDCF of the radio, optical, UV, X-ray, and $\gamma$-ray light curves with respect to the optical $R$-band observations. \textit{From top to bottom:}. $R$--$\gamma$ rays (left), $R$--$1$~keV (right), $R$--$w2$ (left), $R$--$m2$ (right), $R$--$w1$ (left), $R$--$B$ (right), $R$--$V$ (left), $R$--$I$ (right), $R$--$24$~GHz (left), $R$--$8.5$~GHz (right), and $R$--$4.8$~GHz (centre). Different dotted lines denote the contours corresponding to a statistical significance of 1$\sigma$, 2$\sigma,$ and 3$\sigma$.}
    \label{fig:ccf_results_figures}
\end{figure*}

With these considerations, we observe that the optical and UV bands show a high degree of correlation, which is compatible with zero time lag, in agreement with the literature. All the ZDCF curves are represented in Fig.~\ref{fig:ccf_results_figures}. The results of the correlation analysis, with the estimated correlation coefficients, time lags, and significance are reported in Table~\ref{tab:ccf_results}. We note that, due to the poor sampling of the optical filters of \textit{Swift}-UVOT, we only performed the correlation analysis for the three UV light curves. 

The X-ray emission seems to show a delay. However, the sparseness of the data and the large uncertainties due to the faintness of the source in the X-ray domain result in a large uncertainty on the lag estimation, making this delay still compatible with zero. Also, the low level of statistical significance stemming from the these same factors prevents us from extracting more conclusive results for this band.

\begin{table}
\centering
\caption{Results of the inter-band correlation analysis performed for the radio, optical, UV, and $\gamma$-ray datasets with sufficient numbers of observations.}
\label{tab:ccf_results}
\resizebox{\columnwidth}{!}{%
\begin{tabular}{cccc}
\hline
Light curves & $r_{\text{cen}}$ & Time lag $\tau$ [days] & Significance \\ \hline
$R$--$I$ & $0.98 \pm 0.01$ & $10.1 \pm 18.0$ &  >3$\sigma$  \\ \hline
$R$--$V$ & $0.91 \pm 0.01$ & $7.0 \pm 12.4$ &  >3$\sigma$  \\ \hline
$R$--$B$ & $0.92 \pm 0.02$ & $20.1 \pm 29.4$ &  >3$\sigma$  \\ \hline
$R$--$w1$ & $0.93 \pm 0.03$ & $-14.4 \pm 38.7$ &  3$\sigma$  \\ \hline
$R$--$m2$ & $0.90 \pm 0.04$ & $-21.9 \pm 39.2$ &  3$\sigma$  \\ \hline
$R$--$w2$ & $0.93 \pm 0.03$ & $-17.1 \pm 39.8$ &  3$\sigma$  \\ \hline
$R$--$1$~keV & $0.46 \pm 0.12$ & $88.0 \pm 141.5$ &  2$\sigma$  \\ \hline
$R$--4.8 GHz & $-0.61 \pm 0.07$ & $47.7 \pm 54.8$ &  2.5$\sigma$  \\ \hline
$R$--8.5 GHz & $-0.77 \pm 0.05$ & $-25.8 \pm 33.6$ &  2.8$\sigma$  \\ \hline
$R$--24 GHz & $-0.67 \pm 0.15$ & $-22.3 \pm 33.3$ &  2.5$\sigma$  \\ \hline
$R$--$\gamma$ rays & $0.26 \pm 0.12$ & $ -55.4 \pm 248.3 $ &  1.7$\sigma$  \\ \hline
\end{tabular}%
}
\end{table}

We also performed the inter-band correlations between the $R$-band emission and the different radio frequencies. In particular, we looked for possible correlations with the 4.8 GHz, 8.5 GHz, and 24 GHz light curves. Fewer observations are available for the 6.1 GHz and 230 GHz radio bands, and so no correlations can be obtained at these frequencies. We again observe time lags compatible with zero for the 4.8 GHz, 8.5 GHz, and 24 GHz bands. However, in this case, the radio--optical emission seems to be anti-correlated, with significance ranging between 2.5$\sigma$ and 2.8$\sigma$ depending on the frequency. 
{In addition to the anti-correlations observed at zero time lag, the 8.5 GHz and 24 GHz bands also show an increase in the correlation degree at longer delays of $\sim$400~days, albeit with a rather low significance ($\sim$2$\sigma$). This behaviour has typically been observed for the radio emission of blazars in the literature \citep[see e.g.][]{ramakrishnan2016}. However, the lower coverage of the radio data analysed here prevents us from making conclusions related to long-term behaviour, and the anti-correlation observed could also be an effect of variability on timescales inadequately sampled by our analysis.}

\begin{figure}
\centering
        \includegraphics[width=\columnwidth]{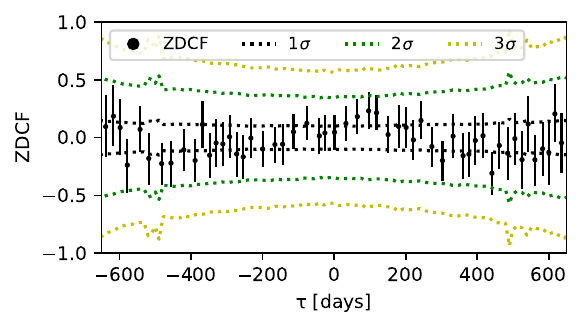}
    \caption{ZDCF between the 8.5 GHz radio and the $\gamma$-ray-band light curves. Different dotted lines denote the contours corresponding to a statistical significance of 1$\sigma$, 2$\sigma,$ and 3$\sigma$.}
    \label{fig:radio_gamma_correlation}
\end{figure}

Finally, for the $\gamma$-ray emission, there is a very faint hint of correlation compatible with zero lag; however, with only a 1.7$\sigma$ significance. Also, the larger bin of the $\gamma$-ray light curve (and therefore lower number of data points) may be affecting the results found here due to a smoothing of the light-curve features. \cite{ramakrishnan2015} did not find significant evidence of a correlation between the radio and $\gamma$-ray emission of 3C~371. We evaluated the 8.5 GHz radio and $\gamma$-ray correlation in the time period considered here, and note the same behaviour reported by these authors. The correlation curve between these two bands is shown in Fig.~\ref{fig:radio_gamma_correlation}.

\section{Jet kinematics}\label{sec4}
\begin{figure}
    \centering
    \includegraphics[width=0.9\linewidth]{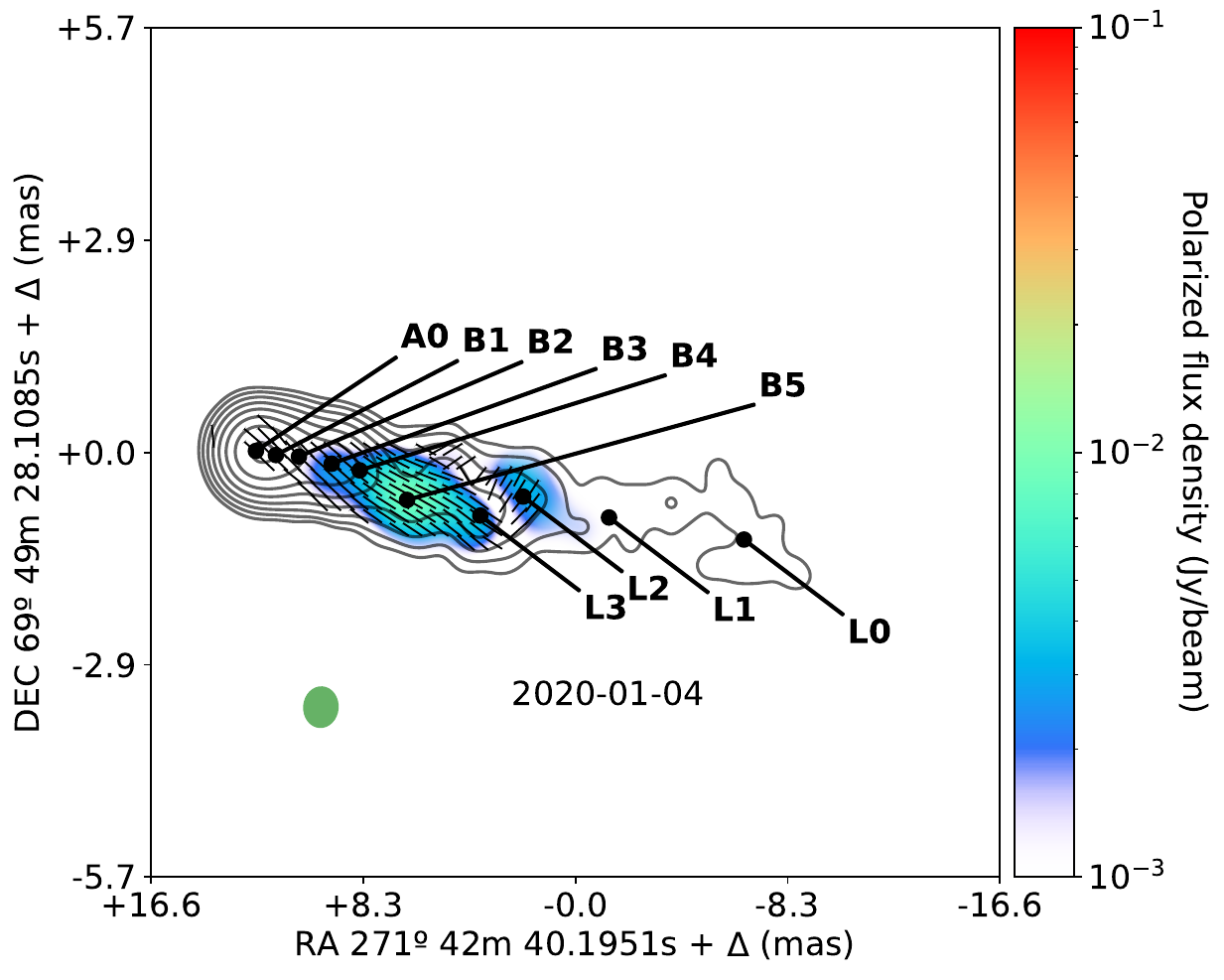}
    \caption{Epoch 2020 Jan 4 VLBI 2 cm image of 3C~371, showing the total flux density (black contours) and the polarised flux density (colour scale). Black line segments overlaid in the image represent the EVPA orientation. The innermost and outermost contours correspond to flux density levels of 5.5 and 0.0015 Jy, with the rest of the contours logarithmically equispaced. The green ellipse represents the beam size. The image highlights the core A0 and all identified components B1-L0.}
    \label{fig:vlbi_epoch_2020-01-04}
\end{figure}

\begin{figure}
    \centering
    \includegraphics[width=0.9\linewidth]{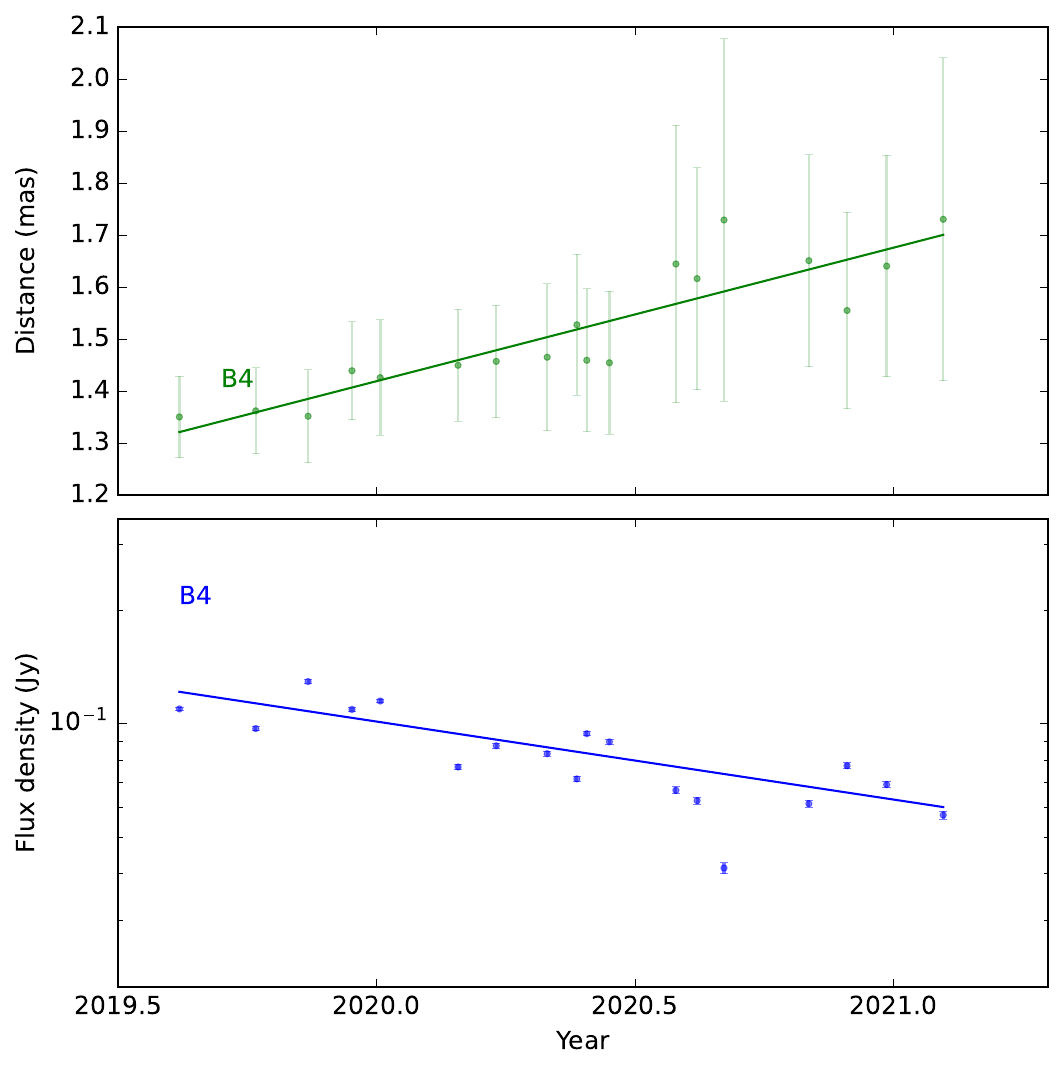}
    \caption{Results of the kinematic analysis for component B4. \textit{Top}: Distance of B4 from the jet core A0 over time. \textit{Bottom}: Flux density of B4 over time.}
    \label{fig:b4_dist_and_flux}
\end{figure}

We performed a kinematic analysis of 2 cm VLBA images of 3C~371 from the MOJAVE monitoring program. 
{The images were first fully calibrated in phase and amplitude by the MOJAVE collaboration. After downloading the calibrated data, we used \texttt{difmap} to produce a VLBA total flux and polarisation image for each of the 18 observing epochs used in this study}. The most prominent features were then fitted to {circular} Gaussian components using \texttt{difmap} and were visually cross-identified along the 18 epochs. {Following the approach presented by \cite{weaver2022}, we aim to characterise the variability timescale of the components by considering that the flux of each knot or component evolves as a decaying exponential function $F=F_0 e^{-t/t_{var}}$. This commonly used assumption allows us to estimate the variability timescale as the time at which the flux of a component decays by a factor $e$.} Among all nine identified components, highlighted in Fig. \ref{fig:vlbi_epoch_2020-01-04}, only one of them (B4) was adequately fitted as a ballistic component with an exponential flux decay (see Fig. \ref{fig:b4_dist_and_flux}), allowing us to extract its speed and timescale of variability. From these estimates, the Doppler factor and apparent speed could be computed \citep{Jorstad:2005} together with the bulk Lorentz Factor and viewing angle. For this component, we obtain an apparent speed of $\beta_{app}=\SI{5.9\pm0.8}{}$ in units of the speed of light, a Doppler factor of $\delta=\SI{6.0\pm1.1}{}$, a bulk Lorentz factor of $\Gamma=\SI{6.0\pm1.8}{}$, and a viewing angle of $\theta=\SI{9.6\pm1.6}{deg}$. {The uncertainties on these parameters were computed through standard error propagation using the uncertainties in the Gaussian model fitting, as given by \texttt{difmap}, and the uncertainties associated to the fits presented in Fig. \ref{fig:b4_dist_and_flux}, following the procedure indicated in \cite{weaver2022}}. An independent preliminary kinematics analysis performed by the MOJAVE team also revealed superluminal motions within the jet with $\beta>4$, which is consistent with the analysis performed here (Kovalev \& Homan, private communication). These results are also compatible with the Doppler factor estimated by \cite{homan2021} ---based on data obtained up to 2019--- by comparing brightness temperatures and proper motions.

Components B1-B3 were also observed to decay in flux; however, their speed is too low to accurately trace their positions and perform the fit. This is in line with previous studies by MOJAVE on this source, where no superluminal motion of the identified components was observed \citep[see e.g.][]{lister2019,lister2021}. Nevertheless, we note that these studies include all MOJAVE data before the period considered here. The variability of this source was rather
low up until 2018, at which point it underwent an unprecedented period of several outbursts and large variations that could be related to the propagation of material along the jet. This propagation of material and acceleration along the jet has been reported for several sources in the past by \citet{homan2015}. Moreover, this brightening and larger variability can also be observed in the VLBI images used here with respect to previous epochs \citep[see][]{lister2019}, with a clear increase in the core flux contemporaneous to the movement observed in the identified components. In addition, the ejection and motion of material along the jet is often observed in blazars after a strong flaring period such as the one observed in 3C~371 between 2018 and 2019 \citep[see e.g.][]{liodakis2020,lico2022,escudero2024}. 

\begin{figure}
    \centering
    \includegraphics[width=\linewidth]{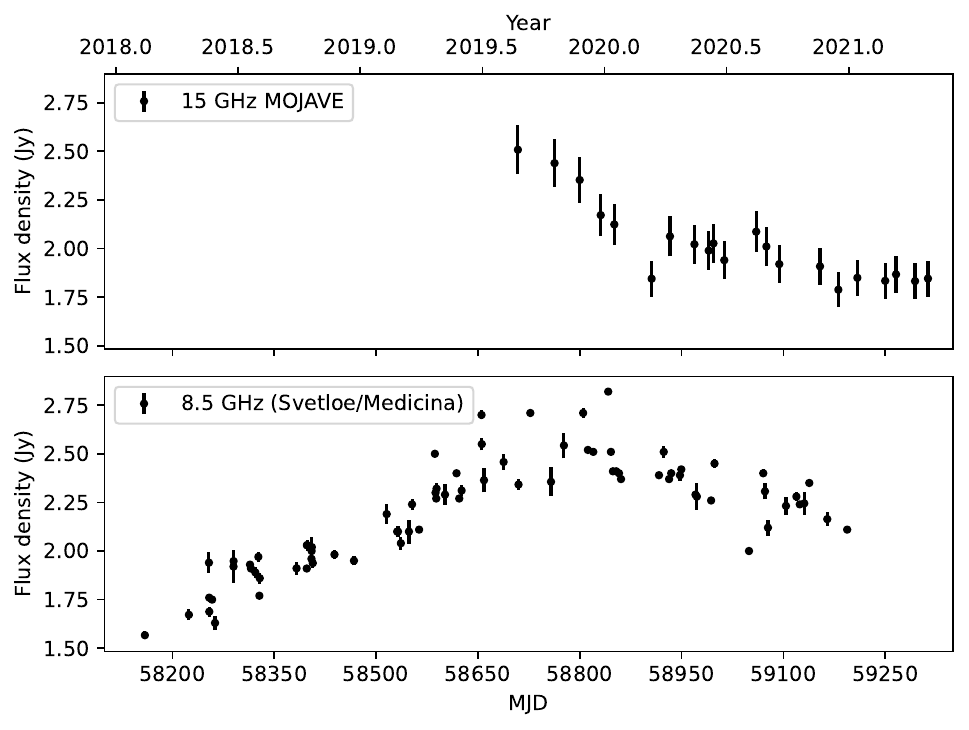}
    \caption{Comparison of radio flux  at 15 GHz and 8.5 GHz. \textit{Top:} Total flux at 15 GHz derived from the MOJAVE VLBI images. \textit{Bottom:} 8.5 GHz radio flux measured by the Svetloe and Medicina radio telescopes.}
    \label{fig:vlbi_total_flux}
\end{figure}

We compare the total 15 GHz flux obtained from the VLBI data analysis shown in Fig.~\ref{fig:vlbi_total_flux} with the radio flux at 8.5 GHz measured by the Svetloe and Medicina radio telescopes. Despite the shorter coverage, the trend followed by the 15 GHz data is correlated with that seen in the other radio bands. We calculate a correlation coefficient between the simultaneous 8.5 GHz and 15 GHz data of $\rho = 0.66$ ($\text{p-value}=0.001$).

Finally, the more external components B5-L0 were clearly observed to move faster from the core. However, their flux does not resemble a decaying exponential, most likely due to the fact that {they are too dim as reflected by their large uncertainties}. Therefore, a reliable kinematic analysis and an estimation of their physical parameters were not possible.

\begin{figure*}
    \subfigure{\includegraphics[width=0.67\columnwidth]{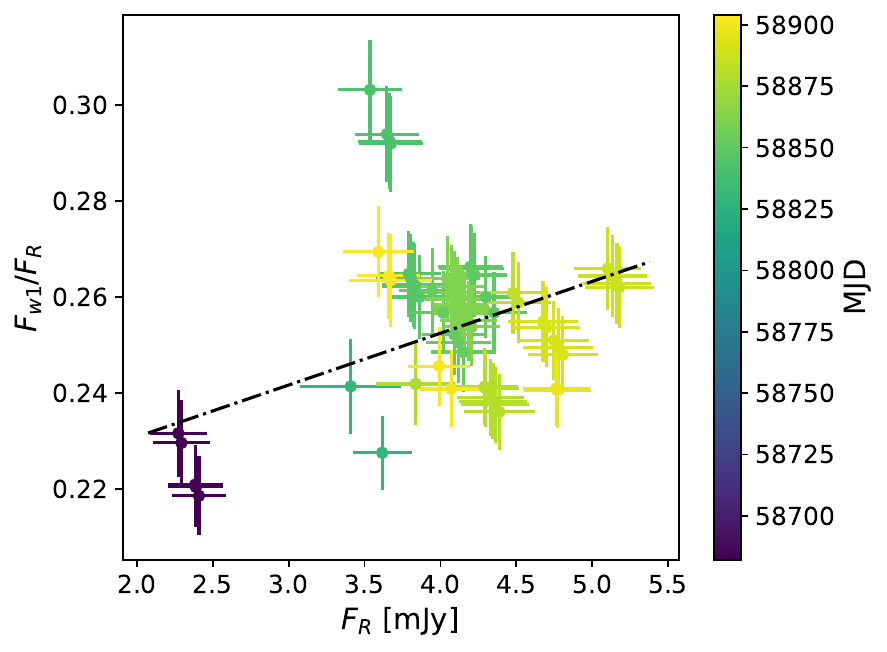}}
    \subfigure{\includegraphics[width=0.67\columnwidth]{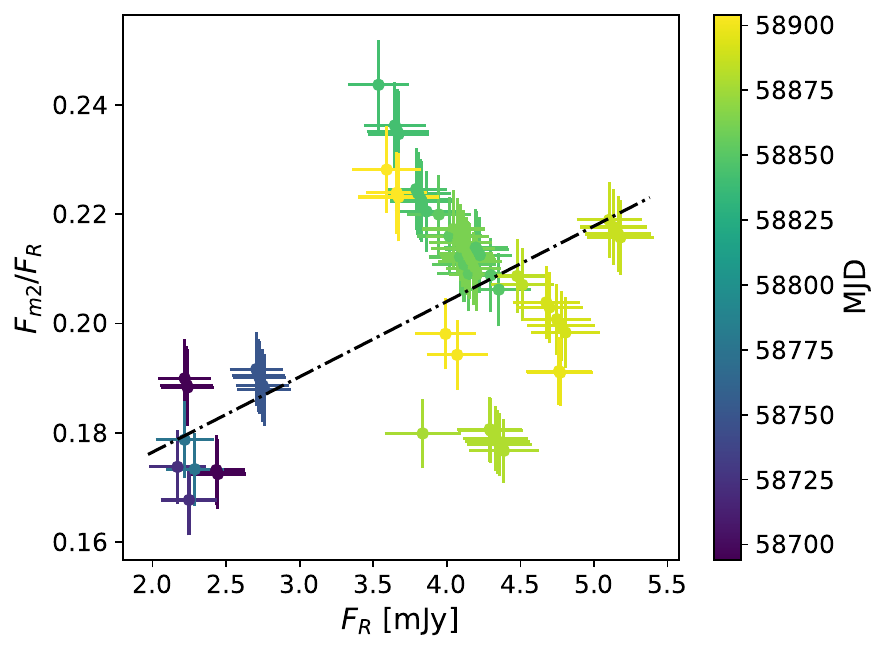}}
    \subfigure{\includegraphics[width=0.67\columnwidth]{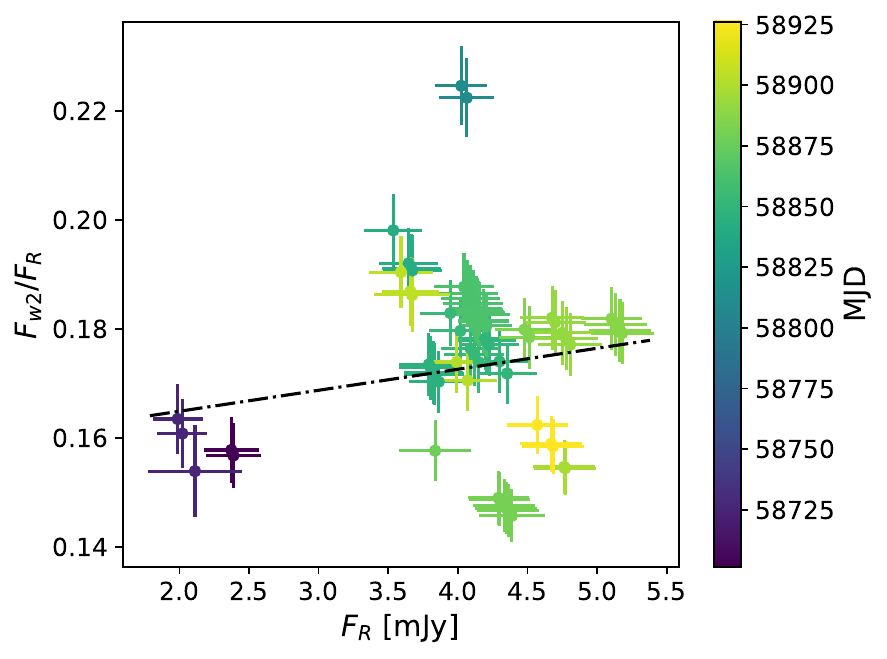}}
    \caption{Dereddened, host-galaxy-corrected colour indices of 3C~371 for the different UV bands with respect to the optical $R$-band flux density. The colour scales represent the MJD of the observations. The black dash-dotted lines correspond to the linear fits describing the long-term colour changes. \textit{Left:} $F_{w1}/F_{R}$ colour index. \textit{Middle:} $F_{m2}/F_{R}$ colour index. \textit{Right:} $F_{w2}/F_{R}$ colour index.} 
    \label{fig:colour_analysis}
\end{figure*}

\section{Spectral variability}\label{sec5}

\subsection{Optical--UV colour variations}

In Paper I, we evaluated the optical colours with respect to the optical $R$ band. These colour relations showed {a mild ---almost achromatic---} long-term bluer-when-brighter (BWB) behaviour, with mixed BWB and redder-when-brighter (RWB) trends on shorter timescales. Here, we extend the analysis to the optical--UV colours between the optical $R$ band and the $w1$, $m2,$ and $w2$ filters. 

In Fig.~\ref{fig:colour_analysis} we present the optical--UV colour relations derived from the data used here. The colour is obtained as the flux ratio between the $R$ band and each UV band, as in Paper I. The temporal coverage provided by \textit{Swift}-UVOT is much shorter than that from the optical $R$-band light curve. However, the long-term evolution seems to follow the mild BWB trend observed between optical bands and with a relatively small colour change, as shown by the low values of the linear fit slopes presented in Table~\ref{tab:colour_analysis_results}. Nevertheless, more data in the UV band are needed to strongly claim a clear long-term colour variability. In addition, as observed from the optical colours, on shorter timescales we can observe a RWB pattern, especially in the $m2-R$ colour relation.

This behaviour follows the results observed in Paper I for the optical colour variability. A very mildly chromatic long-term behaviour is observed, which can be explained as changes of the Doppler factor, possibly caused by orientation changes in the jet that can be ascribed, for instance, to the presence of a helical jet \citep[see e.g.][]{raiteri2021,raiteri2021b}. On the other hand, a variety of colour trends on shorter timescales can be associated to intrinsic jet processes linked to the low-energy synchrotron emission of blazars \citep{raiteri2021,raiteri2021b}. Another possibility involves different relative contributions between the synchrotron emission of the jet and that from the disc. In particular, the RWB trend clearly visible between the $R$ and $m2$ bands may reveal the presence of a faint accretion disc, as this behaviour can be observed during a low flux period \citep{gu2006,gu2011,isler2017,otero-santos2022}. The $m2$ band is also the one showing the highest colour variation of the three UV wavelengths, as seen from the coefficients of the linear fits, although all three remain compatible within the errors. In such cases, the RWB colour evolution could be explained as a relatively stable disc contribution that peaks in the UV band and becomes visible when the synchrotron continuum is faint enough that it no longer completely hinders this thermal contribution, therefore producing a RWB trend until the synchrotron radiation contribution outshines that from the disc at such frequencies \citep[see e.g.][]{villata2006,isler2017}. {However, this scenario is disfavoured given the BL Lac nature of the source; indeed this blazar type typically has inefficient, faint accretion discs that are highly subdominant with respect to the synchrotron emission.} 

{Another possibility is that there are two physical components, a flaring component and a quiescent component, or an underlying flow and a propagating disturbance along the jet. {The} variability introduced by such a propagating component could {manifest on shorter timescales} ---leading to the flaring period that can be observed in the optical light curve--- and would be highly chromatic, {while that} from the underlying flow is expected to be smaller and slower, and might be driven by Doppler factor changes, producing the observed almost achromatic long-term behaviour. This scenario would be consistent with the propagation of component B4 identified in the VLBI data analysis, as the period during which this component is identified is approximately coincident with the strong RWB trend observed in the colour analysis.}

\begin{table}
\centering
\caption{Results of the linear fits to the colour index long-term evolution for the different UV bands with respect to the optical $R$ band.}
\label{tab:colour_analysis_results}
\begin{tabular}{cccc}
\hline
\multirow{2}{*}{Colour} & \multirow{2}{*}{$N$} & \multicolumn{2}{c}{Linear fit $y=a+b \cdot x$}  \\ \cline{3-4} 
  &     &    $a$     &     $b$     \\ \hline
$w1-R$   &  49  &    $0.209 \pm 0.012$      &      $0.010 \pm 0.003$     \\ \hline
$m2-R$   & 59   &    $0.149 \pm 0.010$      &      $0.014 \pm 0.003$     \\ \hline
$w2-R$   &  58  &    $0.157 \pm 0.013$      &     $0.004 \pm 0.003$   \\ \hline
\end{tabular}
\end{table}

\subsection{Optical spectral variability}
\begin{figure}
        \centering
        \includegraphics[width=0.93\columnwidth]{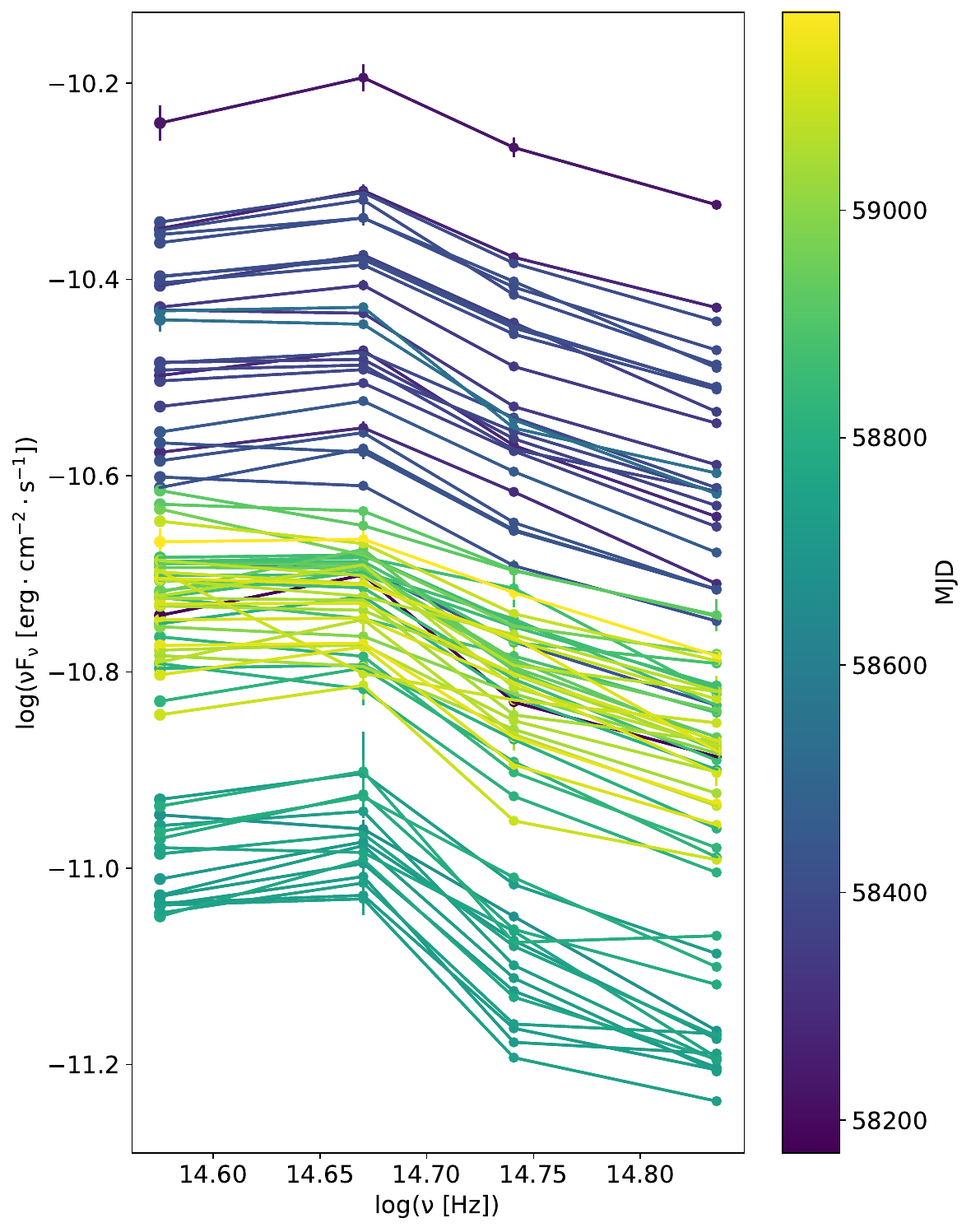}
    \caption{Optical SEDs constructed with quasi-simultaneous $BVRI$ optical data. The colour bar indicates the MJD of the corresponding SED.}
    \label{fig:optical_SEDs}
\end{figure}

Apart from the optical--UV colour variation analysis between the different optical bands performed here and in Paper I, we also constructed {a set of optical spectra} with our entire simultaneous data set in order to test its spectral shape and variability over the analysed period. To this end, we converted the flux densities in milliJanksys (mJy) to units of SED ($\nu F_{\nu}$; i.e. erg~cm$^{-2}$~s$^{-1}$). We constructed a series of SEDs, binning the optical data taken within an interval of 1~day, allowing us to model the variability of the optical synchrotron emission over {timescales of days}, producing a total of 82 SEDs. The constructed SEDs are represented in Fig. \ref{fig:optical_SEDs}.

\begin{figure}
        \includegraphics[width=\columnwidth]{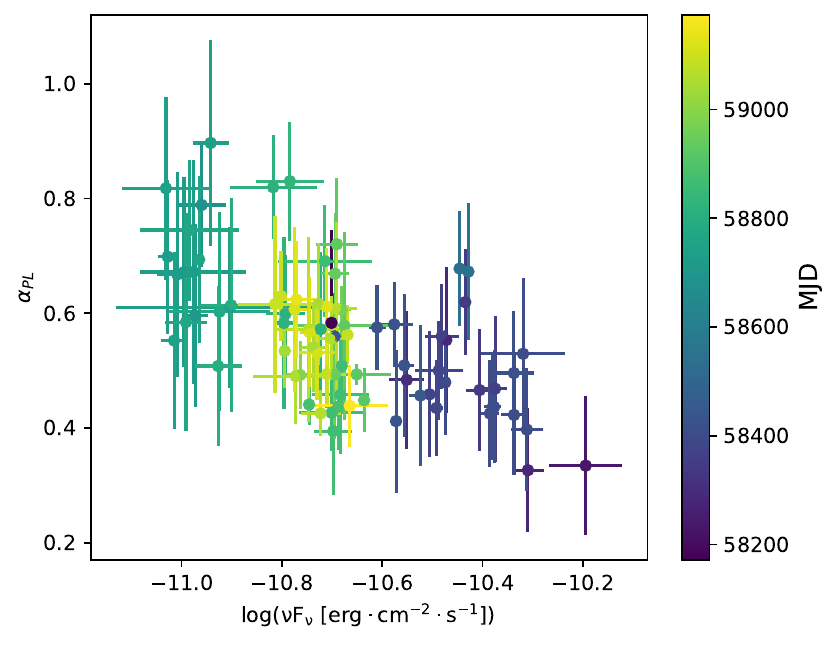}
    \caption{Optical $R$-band flux with respect to the estimated spectral index using a power-law model. The colour bar indicates the MJD of the corresponding SED.}
    \label{fig:optical_flux_index}
\end{figure}

The SEDs show that flux varies by almost one order of magnitude. Upon visual inspection, we observe that the frequency of the synchrotron peak is located in the red/near-infrared part of the spectrum, in agreement with the values of $\nu_{sync}=4.9 \times 10^{14}$~Hz and $\nu_{sync}=7.7 \times 10^{13}$~Hz  reported in the third and fourth catalogues of active galactic nuclei constructed based on data provided by the \textit{Fermi}-LAT telescope \citep{ackermann2015,ajello2020}. We tested a power-law spectral model in the form of $\nu F_{\nu} = K \cdot \nu^{-\alpha}$ to evaluate whether or not there is any relation between the optical spectral index and the flux of the source. We find that the fitted spectral index varies within $\alpha_{PL}=0.33 - 0.90$. Moreover, despite the sometimes large errors in the estimate of the spectral index, we derived a correlation coefficient of $\rho = -0.58$ and a $\text{p-value} = 10^{-8}$, indicating a harder-when-brighter evolution of the optical spectrum with a rather small change of the spectral index, which is therefore compatible with the mild BWB trend found in the optical colour analysis. This pattern can be observed in Fig.~\ref{fig:optical_flux_index}. Owing to the reduced spectral range used for this analysis, we did not consider other, more complex spectral models (e.g. log-parabola function). \cite{xilouris2006} also reported a hardening of the spectrum with increasing flux variations (and the opposite for decreasing flux changes) for 3C~371 during different nights, which is consistent with the results reported here. 

\subsection{X-ray spectral variability}
We also investigated possible X-ray spectral variability using the \textit{Swift}-XRT data. We fit the X-ray spectral index light curve presented in Fig.~\ref{fig:xrt_lcs} to a constant value to decipher whether or not the source displays significant spectral change over the observing period, and evaluated the fit using a $\chi^{2}$ test. The resulting value of $\chi^{2}/\text{d.o.f.}=0.01$ suggests there is no significant variation of the X-ray spectral index and that the light curve is fully compatible with being constant over the period analysed. The fitted constant X-ray spectral index is $\Gamma=1.63 \pm 0.19$. This is also observable from the relation between the X-ray spectral index and the X-ray flux at an energy of 1~keV, shown in the bottom panel of Fig.~\ref{fig:xrt_lcs}. 
Several blazars have shown correlated behaviours between the X-ray flux and spectral index  in past studies, with harder-when-brighter trends in some cases \citep[see e.g.][]{aleksic2013,wang2018,magic2023}. However, even if present in this case, the large errors of the X-ray spectral index of 3C~371 prevent us from drawing any strong conclusions in this regard, as the spectral index reported here is compatible with a constant value.


\subsection{$\gamma$-ray spectral variability}\label{sec:5.4}
We  evaluated the $\gamma$-ray spectral variability using the 15 day binned \textit{Fermi}-LAT light curves. We model the spectrum in each 15 day light curve bin with a power-law function, as shown in Fig.~\ref{fig:fermi_lcs}. This power-law model is consistent with a constant spectral index of $\alpha=2.31 \pm 0.03$ ($\chi^{2}/\text{d.o.f.}=1.02$) over the period considered here. This spectral index is in agreement with the value of $\alpha_{\text{4FGL}}=2.32 \pm 0.02$ reported in the 4FGL-DR4 catalogue after averaging 14 years of LAT data. Therefore, we observe that the $\gamma$-ray spectral index is compatible with a constant and rather steep spectrum, and no significant spectral variability in the $\gamma$-ray band is detected on timescales of 15 days. 

Moreover, we compute the average SED using the 3 years of \textit{Fermi}-LAT data as specified in Sect.~\ref{sec2.3}. Owing to the much larger integration with respect to the 15 day light-curve bins, we test a power law and a log-parabola spectral shape, calculating the possible preference of the log-parabola over the power-law model through a likelihood ratio, as
\begin{equation}
TS_{\text{log-parabola}}=-2\left( \ln{\mathcal{L}_{\text{power law}}}-\ln{\mathcal{L}_{\text{log-parabola}}} \right),
\end{equation}
where $\mathcal{L}$ is the likelihood of the power law and log-parabola in each respective case. The parameters of both spectral models are reported in Table~\ref{tab:fermi_SEDs}. The spectral parameters of both the power-law and the log-parabola models are consistent with those reported in the 4FGL-DR4 catalogue: $\alpha_{\text{4FGL}}=2.32 \pm 0.02$ for the power law; $\alpha_{\text{4FGL}}=2.27 \pm 0.02$ and $\beta_{\text{4FGL}}=0.048 \pm 0.010$ for the log-parabola. For a three-year integration between 2018 and 2020, we only see a 2.53$\sigma$ preference for the log-parabola curvature over the simpler power-law model. This result is nevertheless not surprising, as the 4FGL-DR4 catalogue has a 5$\sigma$ preference for a curved spectrum only after 14 years of integration, which amounts to more than four times the amount of data used in this work.

Moreover, we compute the $\gamma$-ray spectrum for the highest and lowest emission during the period coincident with \textit{Swift} observations, which is the time interval with complete MWL coverage used in Sect.~\ref{sec7} for a broadband interpretation of the emission. For these models, we used time windows of 60 and 90 days for the high and low states, respectively. These integration intervals were chosen in order to obtain a well-characterised SED, as shorter time intervals led to spectra and SEDs with only upper limits. The results for both spectral models are again reported in Table~\ref{tab:fermi_SEDs}. Owing to the short time intervals considered, the use of a power-law shape instead of the log-parabola is justified, as no significant curvature is expected. {This is proven by the spectral analysis performed here, that shows a hint of a curved spectrum observed} after integrating 3 years of data.

\begin{table*}
\centering
\caption{Results of the $\gamma$-ray \textit{Fermi}-LAT spectral fit using power-law and log-parabola spectral shapes, averaging the complete 3~ years of data, and a power law during the time period simultaneous to \textit{Swift} observations for the high and low emission states.}
\label{tab:fermi_SEDs}
\begin{tabular}{ccccccc}
\hline
Time interval & \multirow{2}{*}{Spectral model} & $f_0$ & $E_0$ & \multirow{2}{*}{$\alpha$} & \multirow{2}{*}{$\beta$} & \multirow{2}{*}{$TS_{\text{log-parabola}}$} \\ 
    $[$MJD$]$         &                                 &  [10$^{-12} \times$~MeV$^{-1}$cm$^{-2}$s$^{-1}$]  &  [MeV]  &       &      &   \\ \hline
     58119--59215$^{a}$                          &   Power law                  &  $5.94 \pm 0.24$  &  788.7  &    $2.30 \pm 0.03$   &  --    & \multirow{2}{*}{6.40} \\ 
     &   Log-parabola                  & $6.40 \pm 0.32$   &  788.7  &   $2.29 \pm 0.04$    &   $0.057 \pm 0.025$   &   \\ \hline
     58891--58951$^{b}$                          &   Power law                    &  $8.40 \pm 1.27$  &  788.7  &    $2.42 \pm 0.12$   &   --   &  -- \\ 
     58682--58936$^{c}$  &   Power law                  &   $2.45 \pm 0.63$ &  788.7  &   $2.14 \pm 0.18$    &   --   &  -- \\ \hline
\end{tabular}
{\flushleft
\vspace{-0.2cm}
Notes: $^{a}$3-year averaged spectrum. $^{b}$High state $\gamma$-ray spectrum during the period of \textit{Swift}-UVOT/XRT observations. $^{c}$Low state $\gamma$-ray spectrum during the period of \textit{Swift}-UVOT/XRT observations. \\
}
\end{table*}

\section{Polarised emission}\label{sec6}

\begin{figure}
        \centering
        \includegraphics[width=0.92\columnwidth]{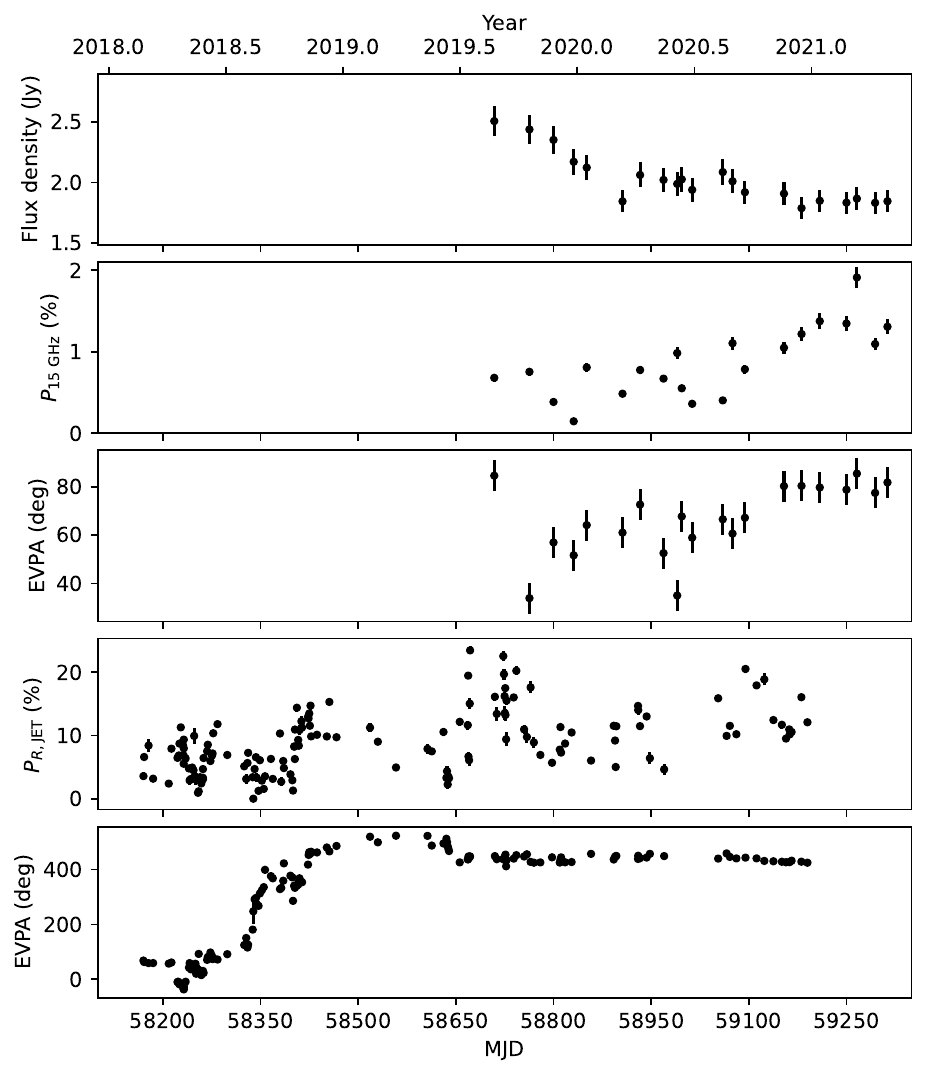}
    \caption{Radio and optical polarisation of 3C~371. \textit{From top to bottom:} 15 GHz flux density, polarisation degree at 15 GHz, EVPA at 15 GHz, intrinsic optical $R$-band polarisation degree, and optical $R$-band EVPA.}
    \label{fig:radio_polarization}
\end{figure}

This source shows a rather low radio polarisation degree at 15 GHz, with values varying between 0.15\% and 1.91\%, as shown in Fig.~\ref{fig:radio_polarization}. As done in Paper I for the optical emission, we evaluated the correlation between the 15 GHz flux density and the 15 GHz polarisation degree. We find that the radio polarisation degree is anti-correlated with the radio flux density, as reflected by the correlation coefficient of $\rho=-0.52$ ($\text{p-value}=0.016$). Therefore, we observe the same behaviour as identified for the optical emission and observed in the past for other sources \citep[for instance BL~Lacertae or S4~0954+65; see][respectively]{raiteri2013,raiteri2023}, where it was interpreted as a cause of the lower Lorentz factor typically measured for BL~Lac objects with respect to FSRQs \citep{raiteri2013}. 

Regarding the EVPA at radio frequencies, during the first epochs covered by the VLBI data there is a certain scatter of the EVPA values between 85$^{\circ}$ and 34$^{\circ}$, contrary to the stability shown by the EVPA in the optical $R$ band in the same period. However, the last seven epochs show a rather constant evolution of the EVPA, with a value of $\sim$80$^{\circ}$. This value is compatible with the polarisation angle orientation of 430$^{\circ}$ in the optical $R$ band after the slow-rotating feature reported in Paper I, considering the difference introduced by the complete 360$^{\circ}$ rotation ($430^{\circ}=70+2\times 180^{\circ}$), meaning that both the radio and optical EVPA end up aligned in approximately the same direction after the studied period. This delay in the stabilisation of the EVPA in the radio band could also indicate a delay of the features in radio frequencies with respect to other frequencies, as mentioned above, meaning that the emission at optical and radio frequencies comes from different regions of the jet, as already suggested by the results obtained from the correlations. Unfortunately, there is no information available on the radio polarisation during the slow rotation observed in the optical polarised emission with which to evaluate the behaviour in the 15 GHz band during this event.
 
\section{SED modelling}\label{sec7}
Finally, we interpret the MWL emission of 3C~371 within the context of the emission processes occurring in blazar jets, providing information on the physical parameters describing the jet and the particle populations responsible for this emission. In order to compare and interpret the observed variability, we construct the SED at the lowest and highest states in which we have complete MWL coverage of the source, that is, during the period in which \textit{Swift}-XRT and \textit{Swift}-UVOT data are available. The high-state SED is produced with the MWL data closest in time to MJD~58921, where the maxima of the $I$- and $R$-band fluxes during this period are observed. On the other hand, the low-state SED is based on data obtained as closely as possible in time to MJD~58726, following the same criteria for the lowest flux observed. The $\gamma$-ray spectrum of each state is obtained by integrating a total of 60 days around the defined MJD for the case of the high state, and 90 days in the case of the low state, as already mentioned in Sect.~\ref{sec:5.4}. 
We corrected the $\gamma$-ray spectra for the absorption introduced by the extragalactic background light (EBL). Even if this effect is small in the energy range of \textit{Fermi}-LAT and especially at such a low redshift, we consider the EBL absorption of the $\gamma$-ray SED using the EBL model from \cite{saldana-lopez2021} in order to model the broadband SED in the most accurate and correct way. 

The modelling was performed using the \texttt{JetSeT}\footnote{\url{https://jetset.readthedocs.io/}} software \citep{tramacere2009,tramacere2011,tramacere2020}. We interpret broadband emission within {a one-zone (blob)} leptonic scenario, where the low-energy emission is produced by synchrotron radiation and the high-energy emission is mainly the product of SSC scattering of low-energy photons, with an electron population described by a power-law function with an exponential cut off with the Lorentz factor, as
\begin{equation}
n(\gamma)=N\gamma^{-p} \exp{\frac{\gamma}{\gamma_{cut}}}, \ \gamma_{min} \leq \gamma \leq \gamma_{max},
\end{equation}
where $\gamma_{min}$, $\gamma_{max}$, and $\gamma_{cut}$ are the minimum, maximum, and cut-off values of the Lorentz factor, $p$ is the spectral index of the electron distribution and $N$ is the normalisation. Moreover, the emission depends on a set of parameters including the size of the emitting region $R$, the magnetic field $B$, the bulk Lorentz factor $\Gamma,$ and the viewing angle $\theta$ of the jet.

Apart from the emission of the jet, we also consider the presence of a dusty torus (DT), a broad line region (BLR), and an accretion {disc that} can contribute in the frequency range from infrared to UV. We used the measurement of the $H_{\beta}$ broad emission line luminosity of $3 \times 10^{40}$~erg~s$^{-1}$ reported by \cite{torrealba2012}, and considering that the luminosity of this emission line corresponds to approximately 22/556 of the entire BLR \citep[see][]{francis1991,ghisellini2015}, we estimate that the BLR has  a total luminosity of $L_{BLR}\sim8 \times 10^{41}$~erg~s$^{-1}$. Using the approximation that relates disc and BLR luminosity as $L_{disc} \simeq 10 \times L_{BLR}$ \citep{ghisellini2015}, the accretion disc luminosity can be approximated as $L_{disc}\sim8 \times 10^{42}$~erg~s$^{-1}$. We fix the accretion efficiency of the disc and the covering factors of the DT and BLR to standard {assumed} values of $\eta=0.08$, $\tau_{DT}=0.1,$ and $\tau_{BLR}=0.1$ {\citep[see e.g.][]{smith1981,ghisellini2015}}. The radii of the BLR and DT are fixed with respect to the
disc luminosity using the relations available from \cite{kaspi2007} and \cite{cleary2007}, respectively,  as
\begin{equation}
R_{BLR, in}=10^{17}\left (\frac{L_{disc}}{10^{45}} \right)^{0.5}
\label{eq:R_BLR_in}
,\end{equation}
\begin{equation}
R_{BLR, out}=1.1 \times R_{BLR, in}
\label{eq:R_BLR_out}
,\end{equation}
\begin{equation}
R_{DT}=2.5\times 10^{18}\left (\frac{L_{disc}}{10^{45}} \right)^{0.5}.
\label{eq:R_DT}
\end{equation}
The presence of the BLR, DT, and disc also introduces the possibility of having an injection of low-energy photons from an external photon field to the jet that can ultimately give rise to significant high-energy emission via EC scattering. Therefore, in our models, we also consider this process when interpreting the high-energy peak of the SEDs. Moreover, as these components are more visible when the synchrotron emission is fainter due to the lower dominance of the synchrotron radiation over their thermal contribution, we first model the SED of the low state, and use the derived values of the DT and disc temperatures, $T_{DT}$ and $T_{disc}$, as fixed parameters in the modelling of the high-state SED.

{{We also consider a second component designed to model the extended jet radio emission, that is, the emission below the synchrotron self-absorption (SSA) frequency of the blob. We model this emission as a power law with an index $\alpha_{radio}$, starting from $\nu_{SSA}^{radio}$ and with an exponential cut-off at $\nu_{cut}^{radio}$. Below the $\nu_{SSA}^{radio}$, the spectral index is set to 5/2. This component mimics an extended jet, where the window between $\nu_{SSA}^{radio}$ and $\nu_{cut}^{radio}$ is produced by the envelope of the optically thin synchrotron emission from different layers of cooled particles, where the age of the particles increases with the distance from the `blazar region', and the window below $\nu_{SSA}^{radio}$ marks the region dominated by the SSA emission from these layers. We stress that this model is purely phenomenological, but nevertheless the inferred value of $\alpha_{SSA}^{radio}$ could provide interesting information about the magnetic field topology of the jet, if we assume a conical geometry for example. Nevertheless, this is beyond the scope of the present analysis and we aim to investigate this topic in a follow-up analysis}}. As seen from the correlation analysis, the emission at radio wavelengths does not show a significant correlation at zero lag as in the rest of the bands, either indicating a possible long delay or a lack of correlation with the rest of the MWL data. This can be explained, for instance, in terms of a strong self-absorption at radio wavelengths, with this radiation being emitted from a different region of the jet at a larger distance from the central black hole, where the radio opacity is lower \citep[see e.g.][]{max-moerbeck2014,magic2023}. Another alternative is that the radio emission is observed after the emitting region has propagated along the jet and has adiabatically expanded \citep{tramacere2022}.

\begin{figure*}
    \centering
    \subfigure{\includegraphics[width=0.46\textwidth]{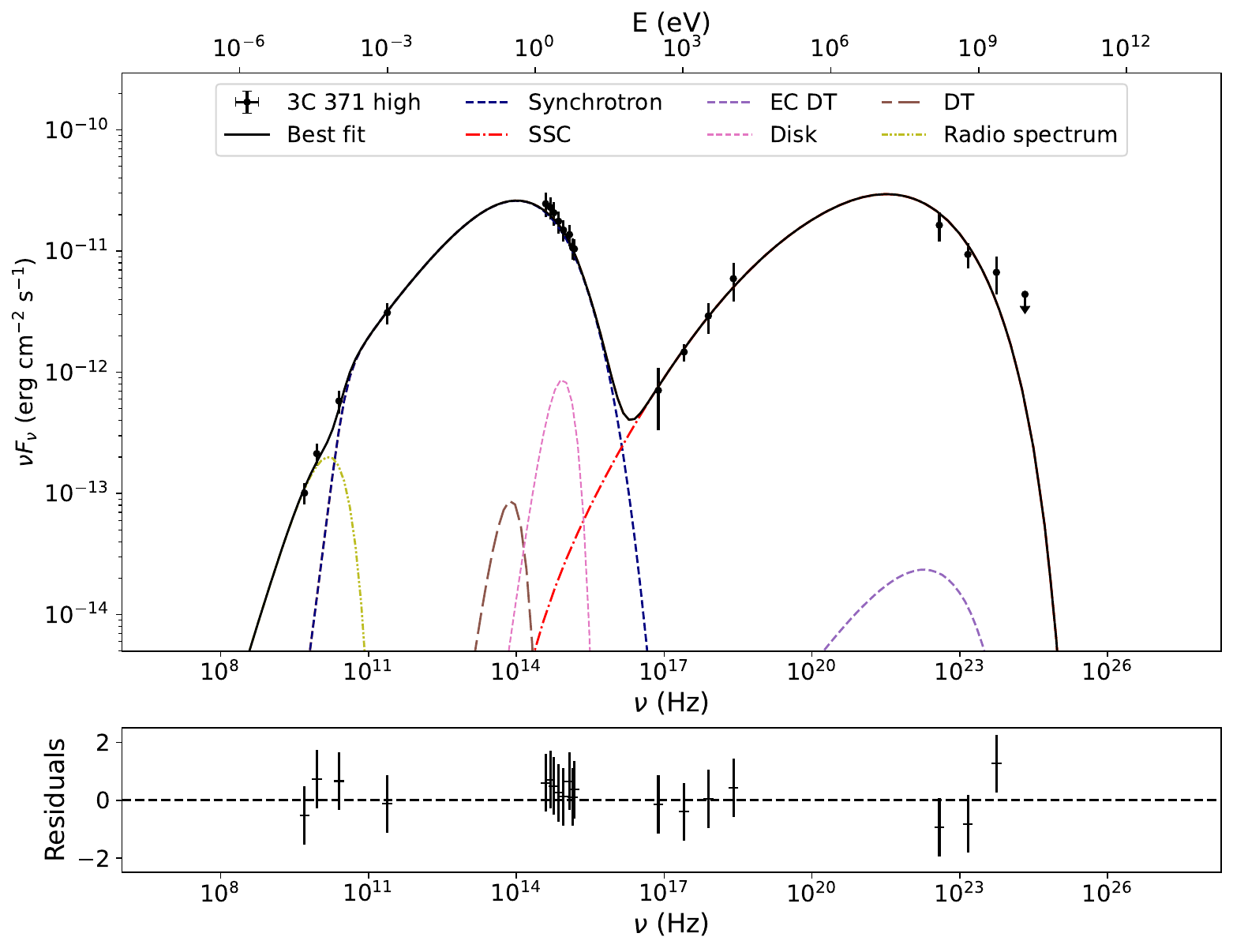}}
    \subfigure{\includegraphics[width=0.46\textwidth]{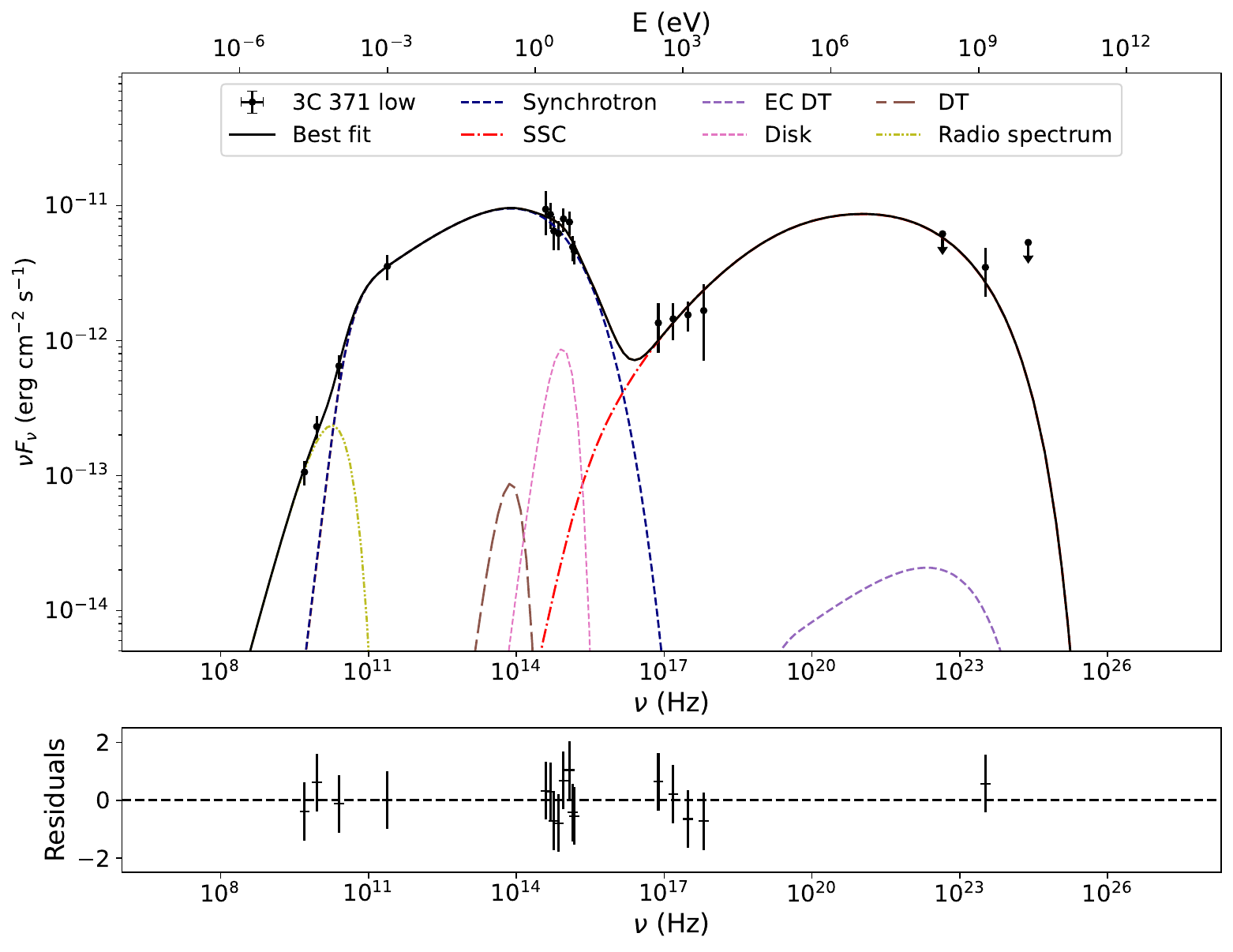}}
    \caption{Broadband SED models for the high- (\textit{left}) and low-brightness states (\textit{right}) of 3C~371. The top panels contain the broadband SEDs, with the contribution of each component and the best-fit model being represented with different markers, lines, and colours, as indicated in the legend. The bottom panels represent the residuals of the fitted models. {We note that the EC scattering contribution via low-energy photons from the BLR  was also considered. However, it is even fainter than that from the DT, and therefore does not fall within the scale of the figures and is not included.}} 
    \label{fig:SED_models}
\end{figure*}

In addition, and in order to simplify and constrain the model and the number of free parameters, we relate the values of $\gamma_{max}$ and $\gamma_{cut}$ through their ratio, allowing values of $\gamma_{cut}/\gamma_{max}$ of between 0.001 and 0.5. The size of the emitting region $R$ is constrained through the opening angle of the jet $\theta_{open}$, {assuming} a value of $5^{\circ}$, as
\begin{equation}
R=R_{H} \times \tan{\theta_{open}},
\label{eq:R}
\end{equation}
where $R_{H}$ is the distance of the region from the central black hole, and is fixed to $10^{18}$~cm. We also restrict the range of variability of the viewing angle and Lorentz factor of the jet, $\theta$ and $\Gamma$, to values around those derived from the VLBI analysis. 
{{For the radio component, we fix the value of $\alpha_{SSA}^{radio}$ from the two radio data points with the lowest frequency to the value of $\alpha_{SSA}^{radio}=-0.35$, and we fix 
$\nu_{SSA}^{radio}=1\times 10^6$ Hz, that is, without loss of generality, below the lowest-frequency radio data point.  
These values were crosschecked with those obtained from the archival data available at the Space Science Data Center (SSDC)\footnote{see \url{https://tools.ssdc.asi.it/SED/}} database and were also found to be relatively constant around the aforementioned values using more than two decades of data. Therefore, we can reliably assume these values to be valid for both models. We then leave the cut frequency $\nu_{cut}$ and $\nu F_{\nu, p}$ as free parameters.}}
Finally, we assume one cold proton per relativistic electron in the jet. {We then performed a model fitting} using the \texttt{JeTSet ModelMinimizer} module with an \texttt{iminuit} minimisation \citep{dembinski2020}. 
The fit is performed considering the components and constraints mentioned above, adding a systematic error of 20\% to the radio, optical, and UV data of the high- and low-state SEDs to avoid biasing the resulting models due to the low statistical uncertainties of these data with respect to the X-ray and $\gamma$-ray spectral points. 
The parameters describing each model are reported in Table~\ref{tab:SED_models}. The two resulting models are shown in Fig.~\ref{fig:SED_models}. A detailed discussion on the reliability of the models and the uncertainties of the parameters is included in Appendix~\ref{sec:appendixC}.

\begin{table}
\begin{center}
\caption{Physical parameters describing the high- and low-state broadband SEDs of 3C~371.}
\label{tab:SED_models}
\resizebox{\columnwidth}{!}{%
\begin{tabular}{lccc}
\hline
            \multicolumn{4}{c}{\it Jet}\\
Parameter & High-state SED & Low-state SED & {Note} \\ \hline
$\gamma_{min}$ & $50.41$ & $120.10$ & {(1)} \\ \hline
$\gamma_{cut}$ & $5.74 \times 10^{3}$ & $1.30 \times 10^{4}$ &  {(1)} \\ \hline
$\gamma_{max}$ & $5.21 \times 10^{5}$ & $4.42 \times 10^{5}$ &  {(1)} \\ \hline
$p$ & $1.85$ & $2.46$ &  {(1)} \\ \hline
$N$ [cm$^{-3}$] & $0.52 \times 10^{2}$ & $0.51 \times 10^{2}$ & {(1)}  \\ \hline
$R$ [cm] & $8.75 \times 10^{16}$ & $8.75 \times 10^{16}$ &  {(4)} \\ \hline
$B$ [G] & $0.20$ & $0.14$ & {(1)}  \\ \hline
$\Gamma$ & $11.47$ & $8.87$ &  {(2)} \\ \hline
$\theta$ [$^{\circ}$] & $10.53$ & $11.09$ & {(2)}  \\ \hline
$L_{disc}$ [erg~s$^{-1}$] & $8 \times 10^{42}$ & $8 \times 10^{42}$ &  {(3)} \\ \hline
$T_{disc}$ [K] & $1.16 \times 10^4$ & $1.16 \times 10^4$ & {(5)}  \\ \hline
$\tau_{BLR}$ & 0.1 & 0.1 & {(6)}  \\ \hline
$R_{BLR, in}$ [cm] & $8.48 \times 10^{15}$ & $8.48 \times 10^{15}$ & {(4)}  \\ \hline
$R_{BLR, out}$ [cm] & $9.33 \times 10^{15}$ & $9.33 \times 10^{15}$ & {(4)}  \\ \hline
$\tau_{DT}$ & 0.1 & 0.1 & {(6)}  \\ \hline
$T_{DT}$ [K] & $10^3$ & $10^3$ &  {(5)} \\ \hline
$R_{DT}$ [cm] & $5.66 \times 10^{17}$ & $5.66 \times 10^{17}$ &  {(4)} \\ \hline
$\theta_{open}$ [deg] & $5$ & $5$ &  {(6)} \\ \hline
\multicolumn{4}{c}{\it Radio component}\\ 
Parameter & High-state SED & Low-state SED &  {Note} \\ \hline
$\alpha_{radio}$ & $-0.35$ & $-0.35$ &  {(3)} \\ \hline
$\nu_{SSA}$ [Hz] & $1.00 \times 10^{6}$ & $1.00 \times 10^{6}$ & {(3)}  \\ \hline
$\nu_{cut}$ [Hz] & $1.15 \times 10^{10}$ & $1.35 \times 10^{10}$ & {(1)}  \\ \hline
$\nu F_{\nu, p}$ [erg~cm$^{-2}$~s$^{-1}$] & $1.90 \times 10^{-13}$ & $2.23 \times 10^{-13}$ &  {(1)} \\ \hline
\end{tabular}
}
\end{center}
{\flushleft
\vspace{-0.2cm}
{Notes: (1) Free parameter. (2) Free parameter with fit range constrained observationally. (3) Parameter fixed observationally. (4) Fixed parameter, functional dependent parameter according to Eqs. (\ref{eq:R_BLR_in}-\ref{eq:R}). (5) Parameter left free in the low-state modelling and fixed for the high-state modelling. (6) Parameter fixed to an assumed value.}
}
\end{table}

The broadband emission is dominated by the non-thermal synchrotron continuum emission, with very faint contributions of the dusty torus and the accretion disc. The accretion disc makes a very minor contribution to the UV emission during the low state, which may support the slight increase in flux in the $w1$ and $m2$ bands when the synchrotron emission is faintest. However, we note that this small bump may also be biased by the fact that the data are not strictly simultaneous due to the poorer time coverage of \textit{Swift}-UVOT. Nevertheless, as expected from BL~Lac objects, this contribution is very small even when the synchrotron emission is at its lowest \citep[see e.g.][]{wang2002,plotkin2012}. Moreover, also as expected from BL~Lac blazars, the contribution of the EC from the torus, BLR, and disc photon fields is also negligible, as observed in Fig.~\ref{fig:SED_models} \citep[e.g.][]{gao2011}. {In fact, the BLR EC contribution is even below the limits of the figure scale, and is therefore not represented in the plots. As a comparison, we included in Appendix~\ref{sec:appendixD} the SED models corresponding to a SSC model, ignoring the EC contribution.} These characteristics indicate that this source is indeed a BL~Lac object, observed in this case under a rather high viewing angle, as derived from both the VLBI analysis ($\theta = 9.6 \pm 1.6$ deg) and the SED models ($\theta \sim 10.5-11$~deg), and as reported in the past by \cite{wrobel1990}. The values derived for $B$ and $\Gamma$ are consistent with those reported by \cite{ghisellini2010} for a larger sample of blazars, combining both BL~Lacs and FSRQs.

The variability between the high- and low-emission states can be explained in terms of changes to the distribution of particles ---mainly through a hardening of the electron distribution during the high state--- as well as a slight increase in the magnetic field $B$ and the bulk Lorentz factor $\Gamma$. This is consistent with the short-timescale variations being caused by intrinsic jet processes ---such as changes of the electron population--- and LTV being attributed to changes of the Lorentz factor. {These changes in the electron population, as discussed above based on the colour analysis results, can be due to an injection of plasma with a harder energy distribution, as indicated by the different colour trend and its coincidence with the detection of B4 in the VLBI analysis. Regarding the Doppler factor changes responsible for the LTV, as previously discussed, there may be more than one possible cause. A plausible mechanism could be related to orientation changes of the jet or the emitting region, either based on, for instance, helical \citep[][]{villata1999,raiteri2004,raiteri2021,raiteri2021b} or precessing jets \cite[see][]{escudero2024b}. These scenarios have been confirmed in a handful of sources \citep[e.g.][]{raiteri2017,cui2023,fuentes2023}. However, Doppler factor variations based on intrinsic Lorentz factor changes of the underlying flow or the component responsible for the LTV could also explain the observed behaviour.} 

 
\begin{table}
\centering
\caption{Energetic report of the SED models derived for the high- and low-emission states.}
\label{tab:SED_energetics}
\begin{tabular}{lcc}
\hline
Parameter & High-state SED & Low-state SED \\ \hline
$U_{e}$ [erg~cm$^{-3}$] & $1.17 \times 10^{-2}$ & $1.33 \times 10^{-2}$ \\ \hline
$U_{B}$ [erg~cm$^{-3}$] & $1.58 \times 10^{-3}$ & $7.46 \times 10^{-4}$ \\ \hline
$P_{rad}$ [erg~s$^{-1}$] & $2.66 \times 10^{44}$ & $5.15 \times 10^{43}$ \\ \hline
$P_{e}$ [erg~s$^{-1}$] & $1.11 \times 10^{45}$ & $7.51 \times 10^{44}$ \\ \hline
$P_{B}$ [erg~s$^{-1}$] & $1.49 \times 10^{44}$ & $4.20 \times 10^{43}$ \\ \hline
$P_{p}$ [erg~s$^{-1}$] & $7.38 \times 10^{45}$ & $4.31 \times 10^{45}$ \\ \hline
$P_{kin}$ [erg~s$^{-1}$] & $8.48 \times 10^{45}$ & $5.06 \times 10^{45}$ \\ \hline
$P_{tot}$ [erg~s$^{-1}$] & $8.90 \times 10^{45}$ & $5.15 \times 10^{45}$ \\ \hline
\end{tabular}
\end{table}

For each model, we also estimated the different contributions to the energy density and jet power, as reported in Table~\ref{tab:SED_energetics}. The estimated magnetic and electron energy densities show a dominance of the latter by a factor of $\sim$10. This is in line with the results reported by \cite{tavecchio2016}, where one-zone SSC models find solutions that are relatively far from an equipartition state between the electron and magnetic energy densities, being in favour of the electron contribution  by a factor of 10-100. The derived contributions to the jet power in forms of electrons ($P_e$), protons ($P_p$), Poynting flux ($P_B$), and radiation ($P_{rad}$) are also consistent with the range of values derived by \cite{ghisellini2010}.

We compared the derived disc luminosity and the power carried by the jet in the form of radiation, $P_{rad}$, with the relation obtained by \cite{ghisellini2014} for these two quantities \citep[see Fig.~1 from][]{ghisellini2014}. The pair of values obtained for the two models presented here are consistent with the relation presented by these latter authors within errors and the 2-3$\sigma$ confidence levels of this relation. The values of 3C~371 can be placed within the diagram from \cite{ghisellini2014} in the same region as other BL~Lac objects with similar faint accretion discs, again supporting the BL~Lac nature of this source.

\section{Summary and conclusions}\label{sec8}
Following the characterisation of the optical emission and variability presented in Paper I, we analysed the MWL emission of 3C~371. To this end, we used the exhaustive monitoring coordinated by the WEBT collaboration in optical and radio wavelengths, as well as UV, X-ray, and $\gamma$-ray data provided by the \textit{Swift}-UVOT/XRT and \textit{Fermi}-LAT telescopes. In addition, we studied the jet kinematics in the considered period through the analysis of VLBI MOJAVE radio data.

\begin{enumerate} 

    \item The $\gamma$-ray band shows variability of  the same amplitude as that of the optical emission. Moreover, the UV and X-ray variability is slightly lower in amplitude than that seen in the optical, but with a much shorter temporal coverage coincident with a relatively faint emitting period. Finally, the radio band presents the least variability among all bands, as commonly observed in $\gamma$-ray-emitting blazars \citep[see e.g.][]{ahnen2017,ahnen2018}.

    \item We find strong correlations with a zero time lag between the optical and UV bands. The X-ray and $\gamma$-ray emission are much more weakly correlated, but with a large uncertainty on the time delays, which are still compatible with zero, and a lower statistical significance. Moreover, the radio emission shows an anti-correlated behaviour at zero time lag, {which could however be an effect of the sampling and coverage of the data}.

    \item We studied the jet kinematics using the MOJAVE VLBI images during the period considered here. Several components are identified, with component B4 showing significant superluminal motion along the jet that allows us to constrain the kinematic parameters, resulting in an estimation of the apparent speed $\beta_{app}$, Doppler factor $\delta$, Lorentz factor $\Gamma,$ and viewing angle $\theta$. The innermost components do not show significant velocities, in agreement with previous MOJAVE analyses \cite[see][]{lister2019,lister2021}. The most external ones on the other hand show significant movement, although they are insufficiently bright for extraction of the kinematic parameters.

    \item The optical--UV colour evolution shows clear mixed BWB and RWB trends on short timescales that can be associated with intrinsic jet processes \citep{raiteri2021,raiteri2021b}. On longer timescales, there appears to be a very mild BWB behaviour similar to that observed between the optical bands. However, the coverage of the UV observations provided by UVOT is much poorer in comparison to that in the optical $R$ band. 

    \item The constructed optical SEDs from the quasi-simultaneous $BVRI$ observations show a variability of approximately one order of magnitude in flux, and present a mild harder-when-brighter change of the spectral index after considering a power-law function to describe the spectrum. The X-ray SEDs and $\gamma$-ray spectral indices were found to be consistent with constant values of $\Gamma=1.63 \pm 0.19$ and $\alpha=2.31 \pm 0.03$, respectively. Moreover, we calculated the average $\gamma$-ray spectrum over the entire period, which shows a $\sim$2.5$\sigma$ preference for a log-parabola spectral shape over a power law.

    \item We characterised the polarised emission at 15 GHz. Similarly to the optical polarisation, the radio polarisation degree is anti-correlated with the radio flux density. Moreover, the radio EVPA shows a rather erratic behaviour during the first studied epochs, stabilising in the latest ones with the same preferred direction as the optical EVPA.

    \item The broadband SED of 3C~371 is successfully modelled within a leptonic scenario, where the bulk of the $\gamma$-ray emission is produced via SSC, with the EC having a negligible impact due to the faintness of the dusty torus and accretion disc photon fields. The variability of the high- and low-state SEDs is explained mainly through a hardening of the electron population distribution.

    \item The derived parameters through the VLBI and broadband SED analyses and the energetics obtained for the relativistic jet are in agreement with 3C~371 being a BL~Lac object observed under a high ($\sim$10~deg) viewing angle.

\end{enumerate}

This thorough characterisation of 3C~371 allows us to interpret its broadband variability and emission, and provides relevant information on the physical parameters describing the jet and confirms its misaligned BL~Lac-type nature \citep{wrobel1990}, which has been debated in the past, with other authors proposing a radio galaxy classification \citep{miller1975}.

\section*{Data availability}
The complete version of Table~\ref{tab:modelfit_results} can be consulted in Zenodo: \url{https://doi.org/10.5281/zenodo.14281786}.

\begin{acknowledgements}
We thank the anonymous referee for the thorough revision of the manuscript and the comments provided to improve the reported results. 
We thank Dr. Yuri Kovalev and Dr. Daniel Homan for the fruitful discussion about the VLBI data analysis and the preliminary independent MOJAVE kinematics results.
J.O.-S., J.E.P. and I.A. acknowledge financial support through the Severo Ochoa grant CEX2021-001131-S funded by MCIN/AEI/ 10.13039/501100011033 and through grants PID2019-107847RB-C44 and PID2022-139117NB-C44. 
J.O.-S. also acknowledges financial support from the projetct ref. AST22\_00001\_9 with founding from the European Union - NextGenerationEU, the \textit{Ministerio de Ciencia, Innovación y Universidades, Plan de Recuperación, Transformación y Resiliencia}, the \textit{Consejería de Universidad, Investigación e Innovación} from the \textit{Junta de Andalucía} and the \textit{Consejo Superior de Investigaciones Científicas}; as well as from INFN Cap. U.1.01.01.01.009.
C.M.R., M.V. and M.I.C. acknowledge financial support from the INAF Fundamental Research Funding Call 2023.
C.M.R. acknowledges support from the Fundación Jesús Serra and the Instituto de Astrofísica de Canarias under the Visiting Researcher Programme 2022-2024 agreed between both institutions.
J.A.-P. acknowledges financial support from the Spanish Ministry of Science and Innovation (MICINN) through the Spanish State Research Agency, under Severo Ochoa Centres of Excellence Programme 2020-2024 (CEX2019-000920-S).
Partly based on data collected by the WEBT collaboration, which are stored in the WEBT archive at the Osservatorio Astrofisico di Torino - INAF (http://www.oato.inaf.it/blazars/webt/); for questions regarding their availability, please contact the WEBT President Massimo Villata ({\tt massimo.villata@inaf.it}).
This article is partly based on observations made with the IAC80 operated on the island of Tenerife by the Instituto de Astrofísica de Canarias in the Spanish Observatorio del Teide. Many thanks are due to the IAC support astronomers and telescope operators for supporting the observations at the IAC80 telescope. This article is also based partly on data obtained with the STELLA robotic telescopes in Tenerife, an AIP facility jointly operated by AIP and IAC.
M.D.J. thanks the Brigham Young University Department of Physics and Astronomy for continued support of the ongoing extragalactic monitoring program at the West Mountain Observatory.
This research was partially supported by the Bulgarian National Science Fund of the Ministry of Education and Science under grant KP-06-H68/4 (2022) and by the Ministry of Education and Science of Bulgaria (support for the Bulgarian National Roadmap for Research Infrastructure).
We acknowledge support by Bulgarian National Science Fund under grant DN18-10/2017 and Bulgarian National Roadmap for Research Infrastructure Project D01-326/04.12.2023 of the  Ministry of Education and Science of the Republic of Bulgaria.
The data presented here were obtained in part with ALFOSC, which is provided by the Instituto de Astrofisica de Andalucia (IAA) under a joint agreement with the University of Copenhagen and NOT.
E.B. acknowledges support from DGAPA-PAPIIT GRANT IN119123.
This work is partly based upon observations carried out at the Observatorio Astronómico Nacional on the Sierra San Pedro Mártir (OAN-SPM), Baja California, Mexico.
G.D., O.V. and M.S. acknowledge support by the Astronomical Station Vidojevica, funding from the Ministry of Science, Technological Development and Innovation of the Republic of Serbia (contract No. 451-03-66/2024-03/200002), by the EC through project BELISSIMA (call FP7-REGPOT-2010-5, No. 256772), the observing and financial grant support from the Institute of Astronomy and Rozhen NAO BAS through the bilateral SANU-BAN joint research project "GAIA astrometry and fast variable astronomical objects", and support by the SANU project F-187.
This work is partly based upon observations carried out at the Hans Haffner Observatory.
The Medicina and Noto radio telescopes are funded by the Ministry of University and Research (MUR) and are operated as National Facility by the National Institute for Astrophysics (INAF). The Sardinia Radio Telescope is funded by the Ministry of University and Research (MUR), Italian Space Agency (ASI), and the Autonomous Region of Sardinia (RAS) and is operated as National Facility by the National Institute for Astrophysics (INAF). We thank the SRT operators Elise Egron, Delphine Perrodin and Mauro Pili for the help with the observations The Submillimeter Array is a joint project between the Smithsonian Astrophysical Observatory and the Academia Sinica Institute of Astronomy and Astrophysics and is funded by the Smithsonian Institution and the Academia Sinica.
We recognise that Maunakea is a culturally important site for the indigenous Hawaiian people; we are privileged to study the cosmos from its summit.
This research has made use of data from the MOJAVE database that is maintained by the MOJAVE team \citep{lister2018}.

\end{acknowledgements}

%
   \bibliographystyle{aa} 
   \bibliography{biblio} 
%
\onecolumn
\begin{appendix} 

\section{Model fitting of the MOJAVE data}\label{sec:appendixA}
\begin{figure}[h]
    \centering
    \includegraphics[width=0.7\linewidth]{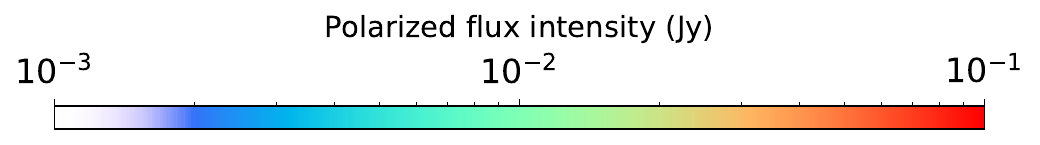}
    \includegraphics[width=0.3\linewidth]{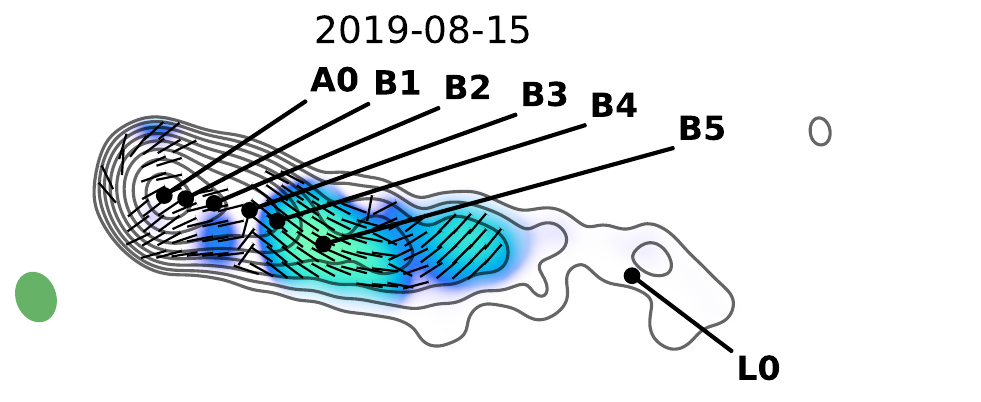}
    \includegraphics[width=0.3\linewidth]{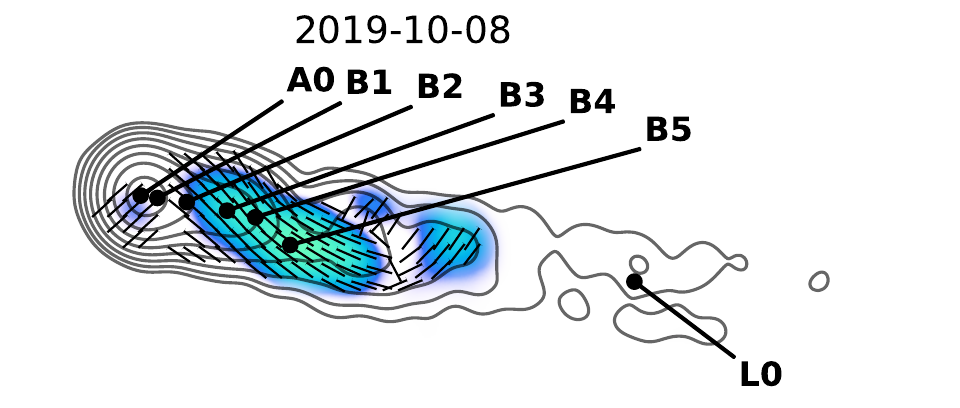}
    \includegraphics[width=0.3\linewidth]{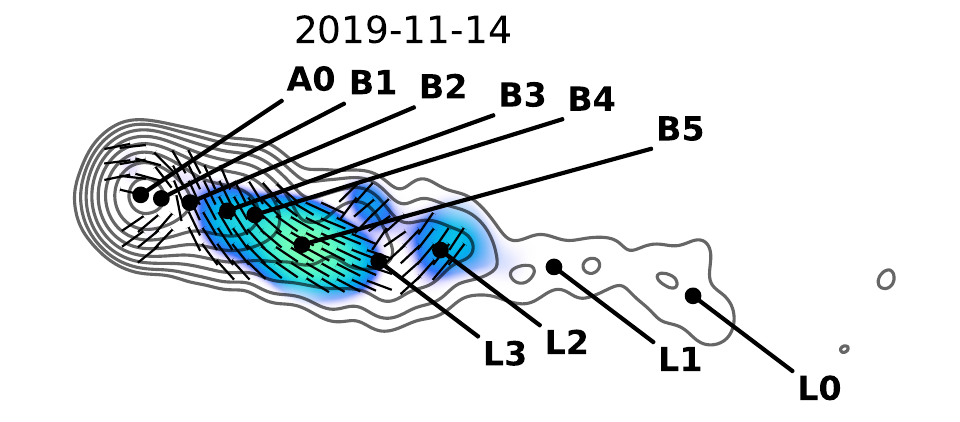}
    \includegraphics[width=0.3\linewidth]{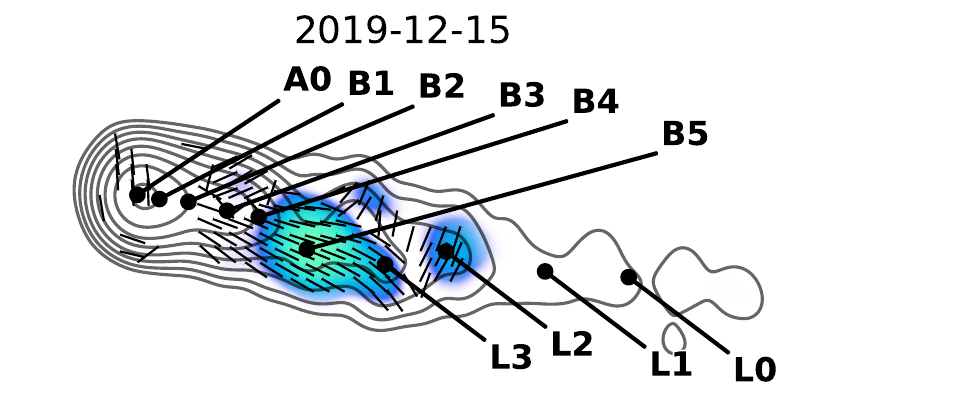}
    \includegraphics[width=0.3\linewidth]{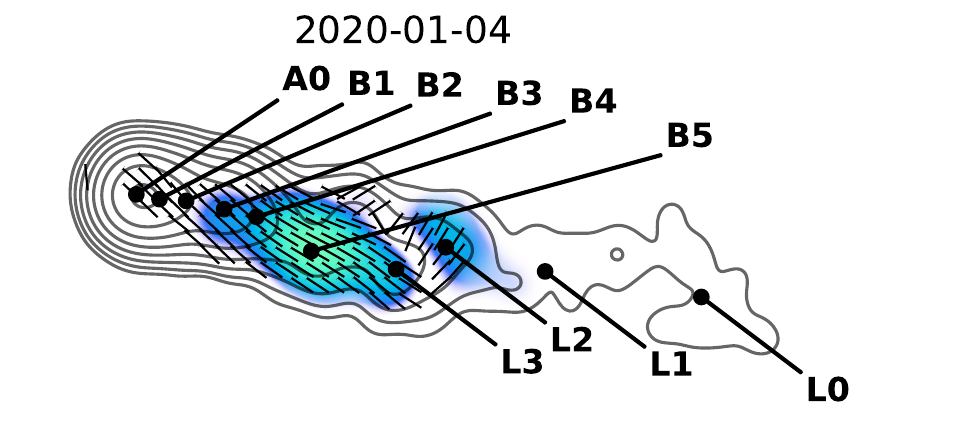}
    \includegraphics[width=0.3\linewidth]{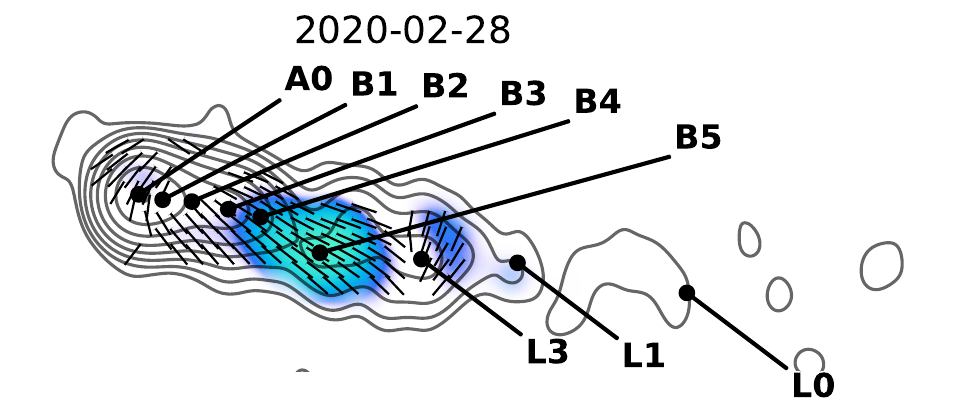}
    \includegraphics[width=0.3\linewidth]{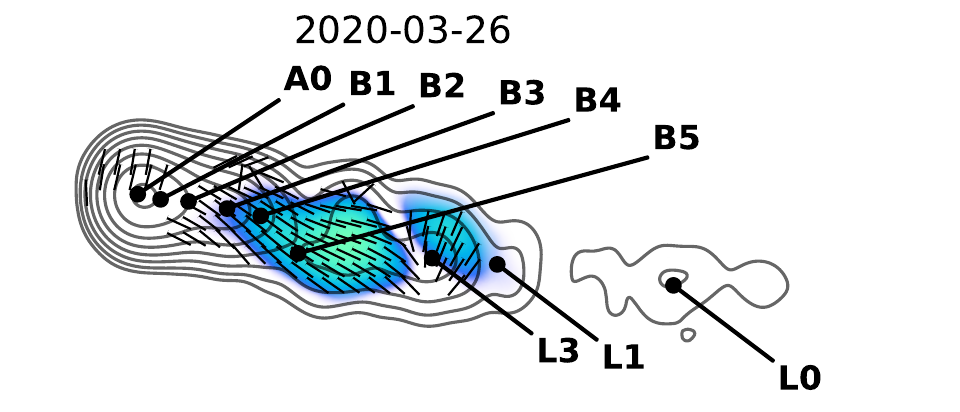}
    \includegraphics[width=0.3\linewidth]{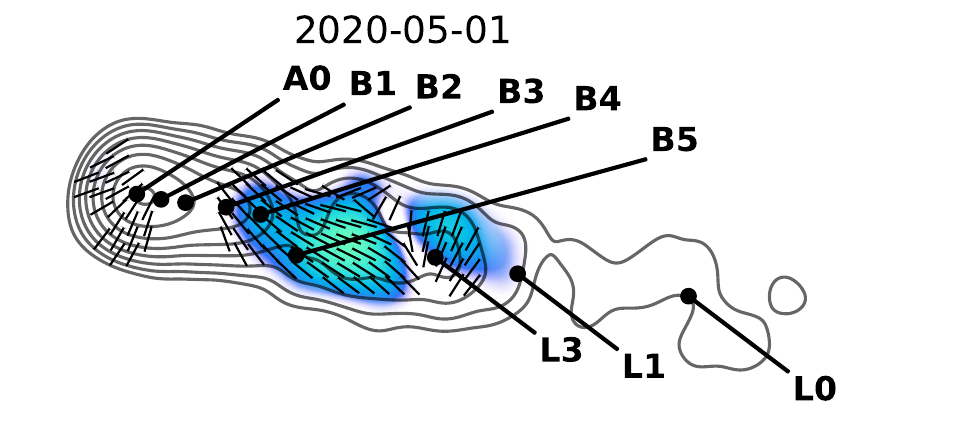}
    \includegraphics[width=0.3\linewidth]{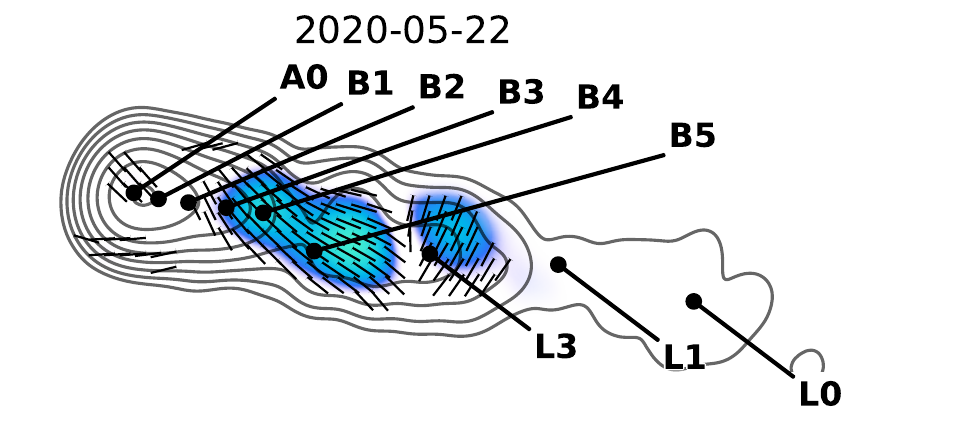}
    \includegraphics[width=0.3\linewidth]{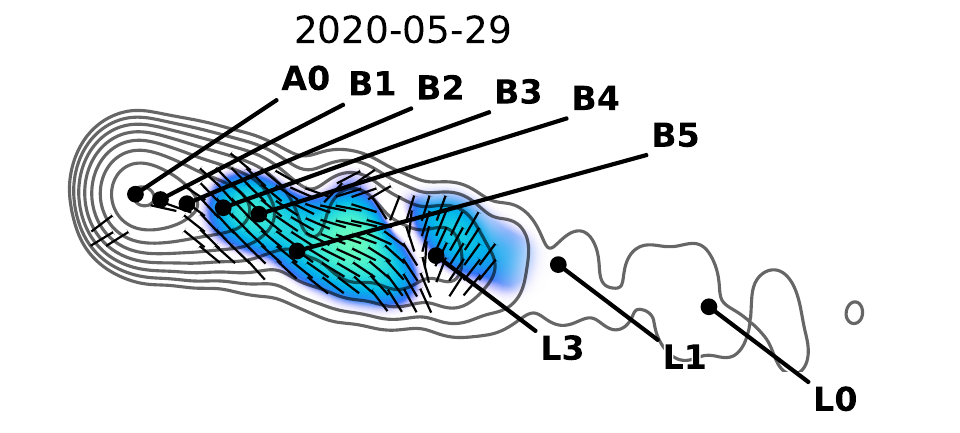}
    \includegraphics[width=0.3\linewidth]{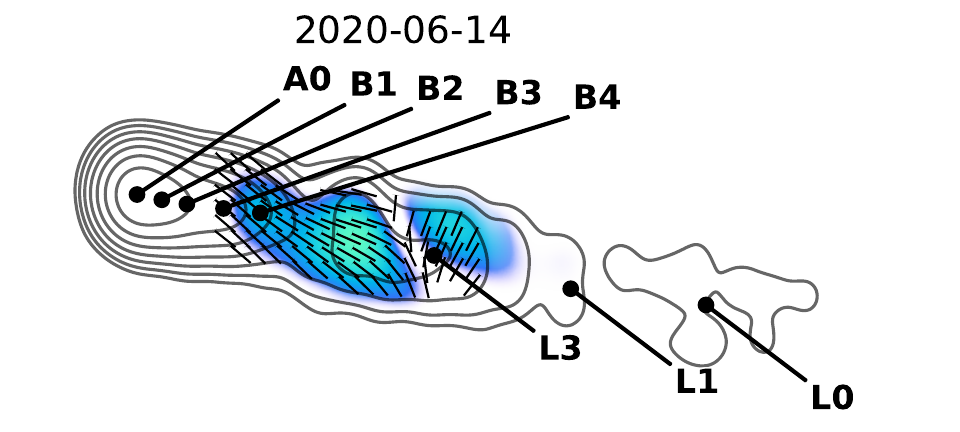}
    \includegraphics[width=0.3\linewidth]{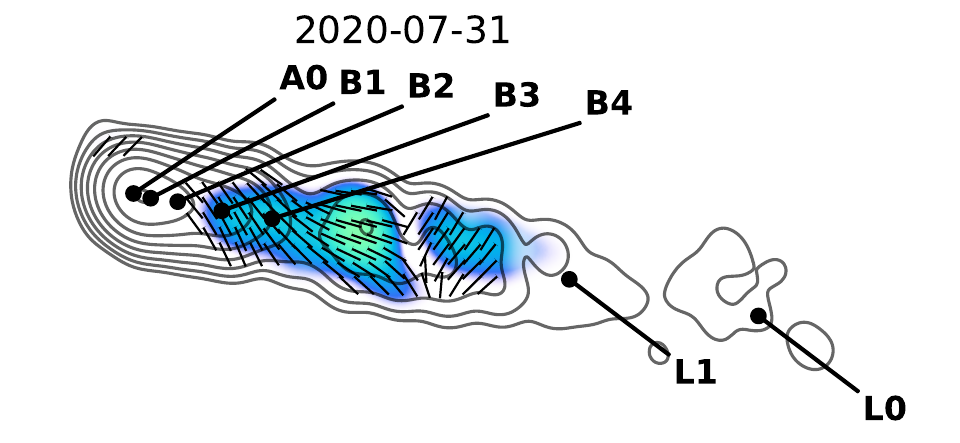}
    \includegraphics[width=0.3\linewidth]{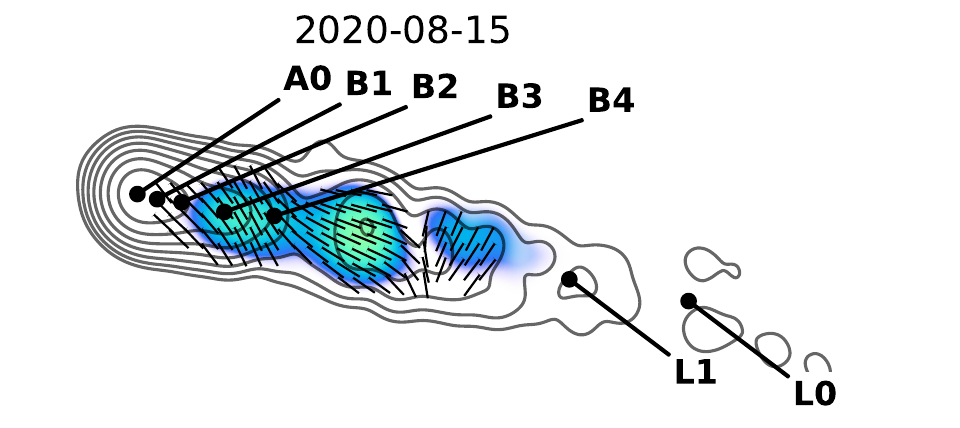}
    \includegraphics[width=0.3\linewidth]{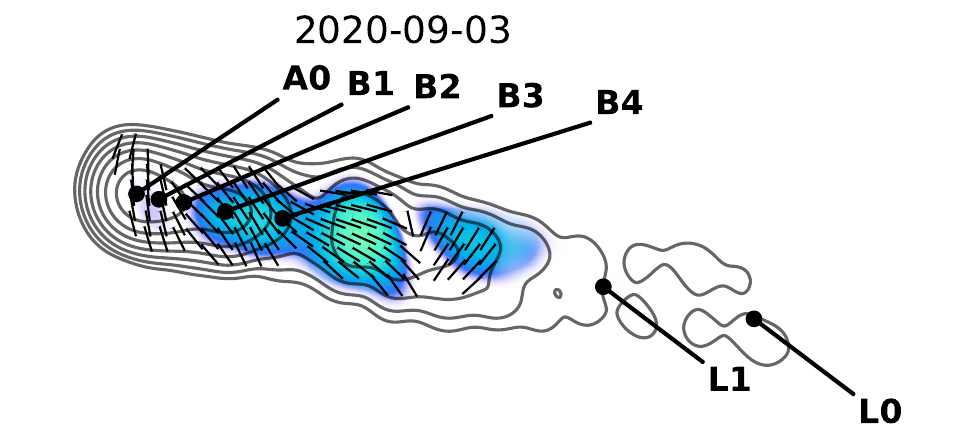}
    \includegraphics[width=0.3\linewidth]{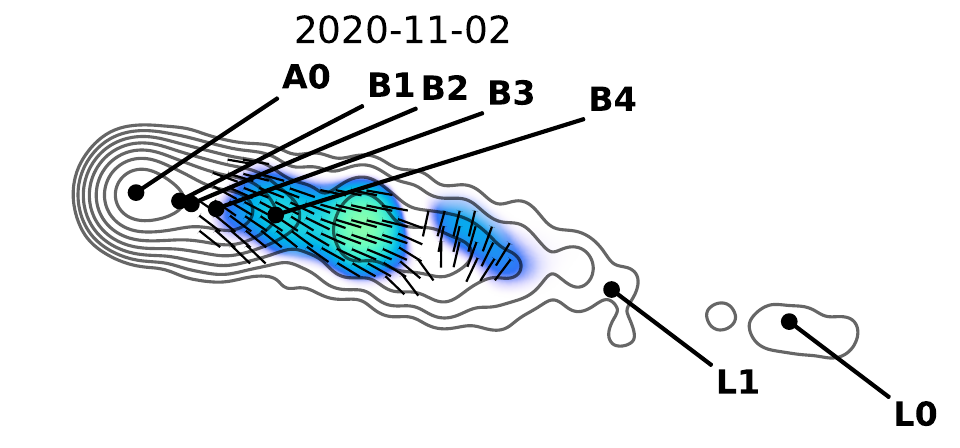}
    \includegraphics[width=0.3\linewidth]{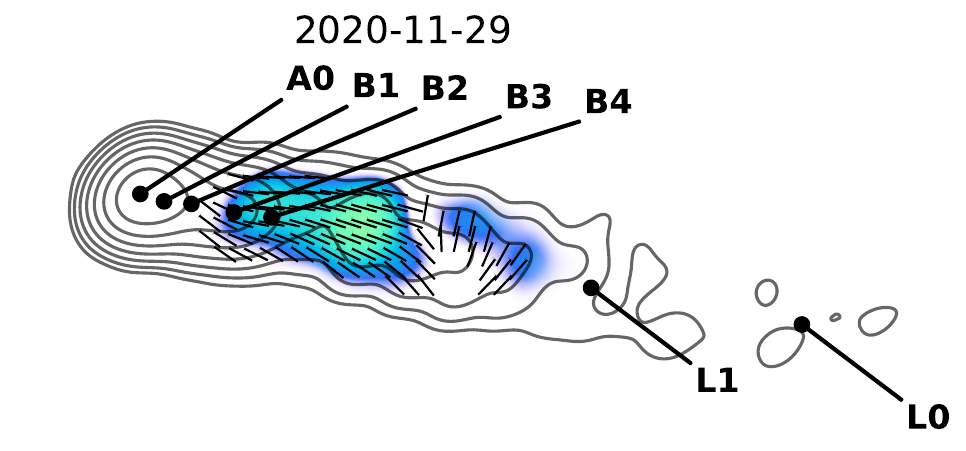}
    \includegraphics[width=0.3\linewidth]{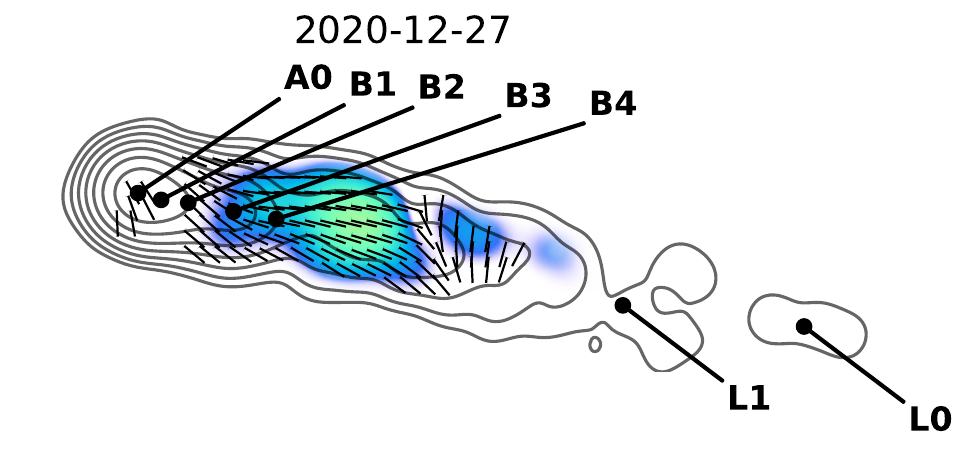}
    \includegraphics[width=0.3\linewidth]{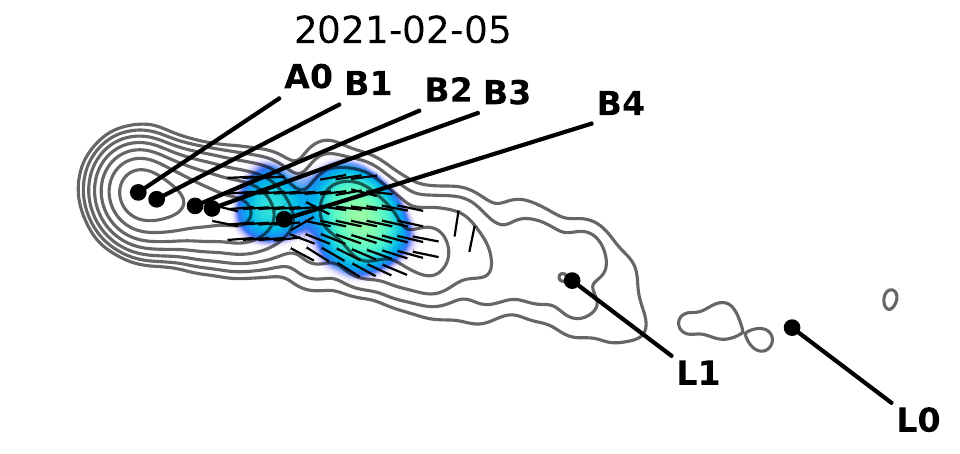}
    \caption{VLBI 2 cm images of 3C~371 for each observing epoch included in the analysis, showing the total flux density (black contours) and the polarised flux density (colour scale). 
    Innermost and outermost contours correspond to flux density levels of 5.5 and 0.0015 Jy, with the rest of the contours logarithmically equispaced.
    The green ellipse at the top left corner (next to the first images) represents the beam size. The images highlight the positions of the core A0 and all identified components B1-L0. The date corresponding to each image is written on top of each plot.}
    \label{fig:VLBI_images_all}
\end{figure}

{Here we include the modelfit results from the VLBI data analysis of all 18 epochs included in this work, from August 15, 2019 to February 5, 2021. Figure~\ref{fig:VLBI_images_all} shows the VLBI images of all epochs, with the identification of the different components reported in the manuscript.}

{Moreover, in Table~\ref{tab:modelfit_results} below we compile the modelfit results obtained from our analysis of the VLBI data during all epochs, for all identified components. This table includes the information of the flux density and the position of each identified component for the different epochs included here. Detailed information on the products of the analysis can be found in the \texttt{difmap} documentation\footnote{\url{http://www.ira.inaf.it/Computing/manuals/difmap/cookbook.ps}}.}

\newpage

\begin{table*}[h]
\begin{center}
\caption{Modelfit results derived from the VLBI data analysis for all epochs and each identified component.}
\label{tab:modelfit_results}
\begin{tabular}{ccccc}
\hline
\multirow{2}{*}{Component} & \multirow{2}{*}{Epoch} & $F$ & X & Y  \\  
  &   & [Jy] & [mas] & [mas]  \\ \hline
A0 & 2019 Aug 14 & 1.30 & -0.056 & 0.007 \\ \hline
B1 & 2019 Aug 14 & 0.41 & 0.193 &  -0.030\\ \hline
B2 & 2019 Aug 14 & 0.33 & 0.525 & -0.084 \\ \hline
 & 2019 Aug 14 & 0.07 & 2.497 & -0.579 \\ \hline
B3 & 2019 Aug 14 & 0.22 & 0.939 & -0.165 \\ \hline
B4 & 2019 Aug 14 & 0.11 & 1.262 & -0.290 \\ \hline
 & 2019 Aug 14 & 0.05 & 3.463 & -0.639 \\ \hline
B5 & 2019 Aug 14 & 0.04 & 1.802 & -0.559 \\ \hline
L0 & 2019 Aug 14 & 0.02 & 5.403 & -0.929 \\ \hline
\end{tabular}
\end{center}
\vspace{-0.2cm}
Notes: Columns correspond to the identified components, the corresponding epoch, the flux density and the position of the component in milli-arcseconds. Components without an associated name correspond to those used for the modelfitting but not included in our analysis due to high noise, making them not suitable for the kinematic evaluation and for tracing their evolution and continuity. More details on the modelfit results can be consulted in the documentation of \texttt{difmap}. The complete table is available in electronic form. The first rows are shown here as an example for guidance regarding its form and content. The complete version of the table is available in Zenodo: \url{https://doi.org/10.5281/zenodo.14281786}. \\
\end{table*}

\section{Radio polarisation data analysis}\label{sec:appendixB}
In this appendix we include the products of the polarisation estimation in the 15-GHz radio band based on the MOJAVE VLBI data for the 18 epochs analysed here. All the estimates were found to be consistent with those reported by the MOJAVE team in the dedicated webpage\footnote{see \url{https://www.cv.nrao.edu/MOJAVE/sourcepages/1807+698.shtml}} to the observations of 3C 371. {Following the typical VLBI radio data analysis approach, we assume a 5\% systematic uncertainty to account for amplitude calibration errors when estimating the total flux and the Stokes parameters \citep[][]{jorstad2017}. The latter is translated into a 7\% error of the polarised flux.} For the error of the polarisation {angle} we assume a conservative value of $\Delta \theta_{sys}=5$~deg {of systematic uncertainty}, consistent with that obtained assuming the 5\%-10\% typically used for VLBI radio images \citep[see for instance][]{jorstad2017,kim2019,weaver2022}\footnote{see also \url{https://science.nrao.edu/facilities/vlba/docs/manuals/oss/amp-cal}}. {This systematic error is later added in quadrature to obtain the total EVPA uncertainty as $\Delta \theta = (\Delta \theta_{sys}^{2} + (\Delta P/P)^{2})^{1/2}$}.

\begin{table*}[h!]
\begin{center}
\caption{Radio polarisation estimates in the 15-GHz band derived from our MOJAVE VLBI data analysis.}
\label{tab:radio_polarization_estimates}
\begin{tabular}{cccccccc}
\hline
\multirow{2}{*}{Date} & \multirow{2}{*}{MJD} & I & $\Delta$I & P & $\Delta$P & $\theta$ & $\Delta \theta$ \\  
  &   & [Jy] & [Jy] & [\%] & [\%] & [deg] & [deg] \\ \hline
2019 Aug 14  & 58709 & 2.51 & 0.13 & 0.68 & 0.05 & 84.5 & 6.41 \\ \hline
2019 Oct 07  & 58763 & 2.44 & 0.12 & 0.75 & 0.05 & 33.9 & 6.41 \\ \hline
2019 Nov 13  & 58800 & 2.35 & 0.12 & 0.38 & 0.03 & 56.9 & 6.41 \\ \hline
2019 Dec 14  & 58831 & 2.17 & 0.11 & 0.15 & 0.01 & 51.6 & 6.41 \\ \hline
2020 Jan 03  & 58851 & 2.12 & 0.11 & 0.81 & 0.06 & 64.1 & 6.41 \\ \hline
2020 Feb 27  & 58906 & 1.84 & 0.09 & 0.48 & 0.03 & 61.0 & 6.41 \\ \hline
2020 Mar 25  & 58933 & 2.06 & 0.10 & 0.78 & 0.05 & 72.6 & 6.41 \\ \hline
2020 Apr 30  & 58969 & 2.02 & 0.10 & 0.67 & 0.05 & 52.2 & 6.41 \\ \hline
2020 May 21  & 58990 & 0.99 & 0.10 & 0.98 & 0.07 & 35.0 & 6.41 \\ \hline
2020 May 28  & 58997 & 2.03 & 0.10 & 0.55 & 0.04 & 67.7 & 6.41 \\ \hline
2020 Jun 13  & 59013 & 1.94 & 0.10 & 0.36 & 0.03 & 58.9 & 6.41 \\ \hline
2020 Jul 30  & 59060 & 2.09 & 0.10 & 0.40 & 0.03 & 66.5 & 6.41 \\ \hline
2020 Aug 14  & 59075 & 2.01 & 0.10 & 1.10 & 0.08 & 60.6 & 6.41 \\ \hline
2020 Sep 02  & 59094 & 1.92 & 0.10 & 0.79 & 0.05 & 67.1 & 6.41 \\ \hline
2020 Nov 01  & 59154 & 1.91 & 0.10 & 1.05 & 0.07 & 80.2 & 6.41 \\ \hline
2020 Nov 28  & 59181 & 1.79 & 0.09 & 1.22 & 0.09 & 80.3 & 6.41 \\ \hline
2020 Dec 26  & 59209 & 1.85 & 0.09 & 1.37 & 1.00 & 79.7 & 6.41 \\ \hline
2021 Feb 05  & 59250 & 1.83 & 0.09 & 1.35 & 0.09 & 78.8 & 6.41 \\ \hline
2021 Feb 21  & 59266 & 1.87 & 0.09 & 1.91 & 0.13 & 85.4 & 6.41 \\ \hline
2021 Mar 21  & 59294 & 1.83 & 0.09 & 1.09 & 0.08 & 77.4 & 6.41 \\ \hline 
2021 Apr 09  & 59313 & 1.85 & 0.09 & 1.31 & 0.09 & 81.7 & 6.41 \\ \hline  
\end{tabular}
\end{center}
\vspace{-0.2cm}
{Notes: A typical value of 5~deg was adopted as the {systematic} uncertainty of the EVPA, {added in quadrature to obtain the total error.}}
\end{table*}

\section{Uncertainty evaluation of the SED parameters}\label{sec:appendixC}
\begin{figure*}[h!]
    \subfigure{\includegraphics[width=0.49\textwidth]{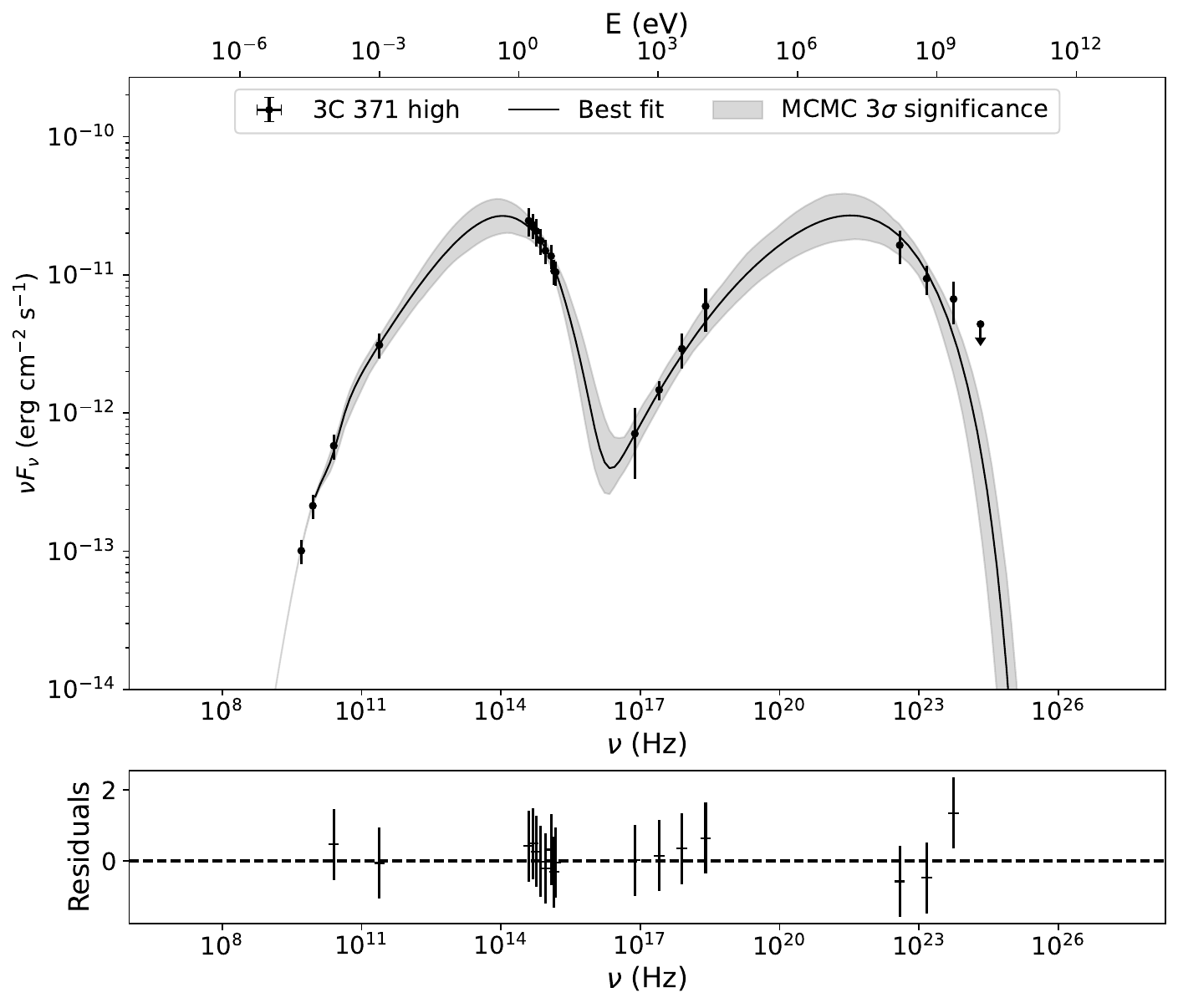}}
    \subfigure{\includegraphics[width=0.49\textwidth]{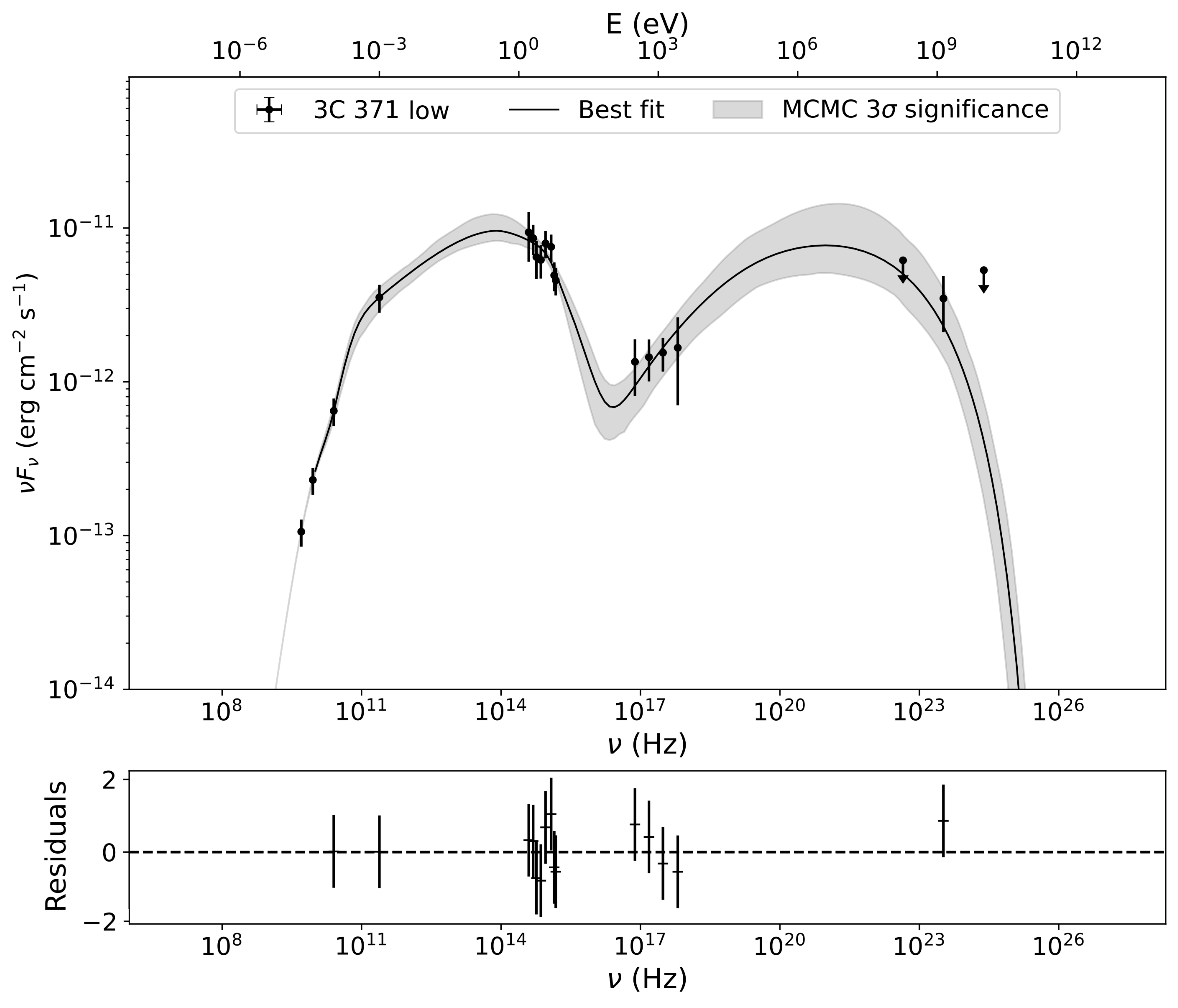}}
    \caption{Broadband SED models of 3C~371. \textit{Left:} High-state SED. \textit{Right:} Low-state SED. The spectral points of each model are represented with the blue markers. Downward pointing arrows represent upper limits. The black solid line corresponds to the best fit and the grey contours represent the 3$\sigma$ confidence region of the fit. The bottom panel of each SED shows the residuals of the model.} 
    \label{fig:SED_models_mc}
\end{figure*}

\begin{figure*}
\includegraphics[width=\textwidth]{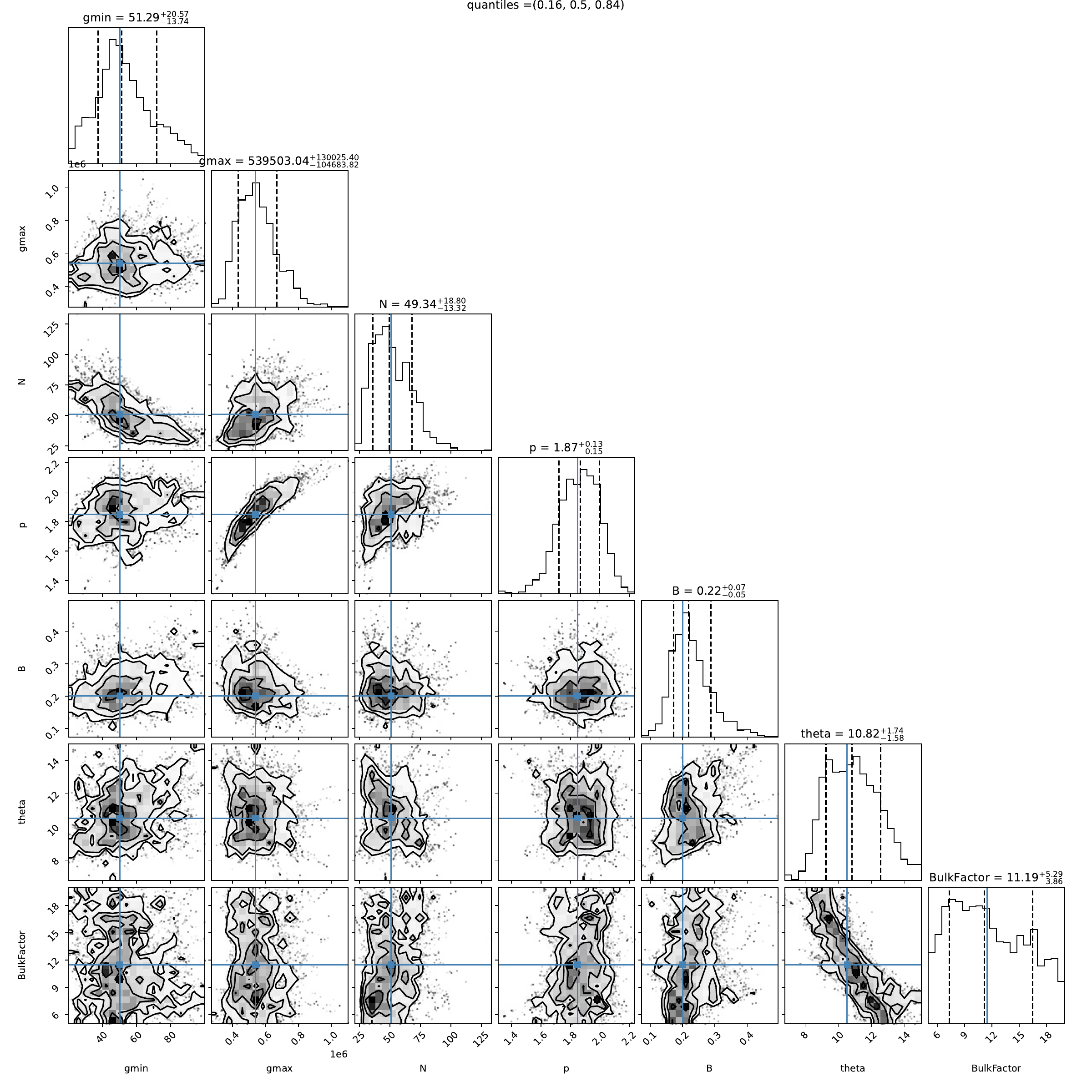}
\caption{Corner plot summary of the parameter distributions estimated through the MCMC procedure for the high state SED model. The panels on the diagonal show the distribution of each parameter, indicating the best fit value and the derived uncertainties with the black solid and dashed lines, respectively. The best fit value of each parameter and its corresponding uncertainties are also indicated on top of each histogram. The panels below the diagonal show the 2-dimensional projections of the histograms for each pair of parameters.} 
\label{fig:SED_model_high_cornerplot}
\end{figure*}

\begin{figure*}
\includegraphics[width=\textwidth]{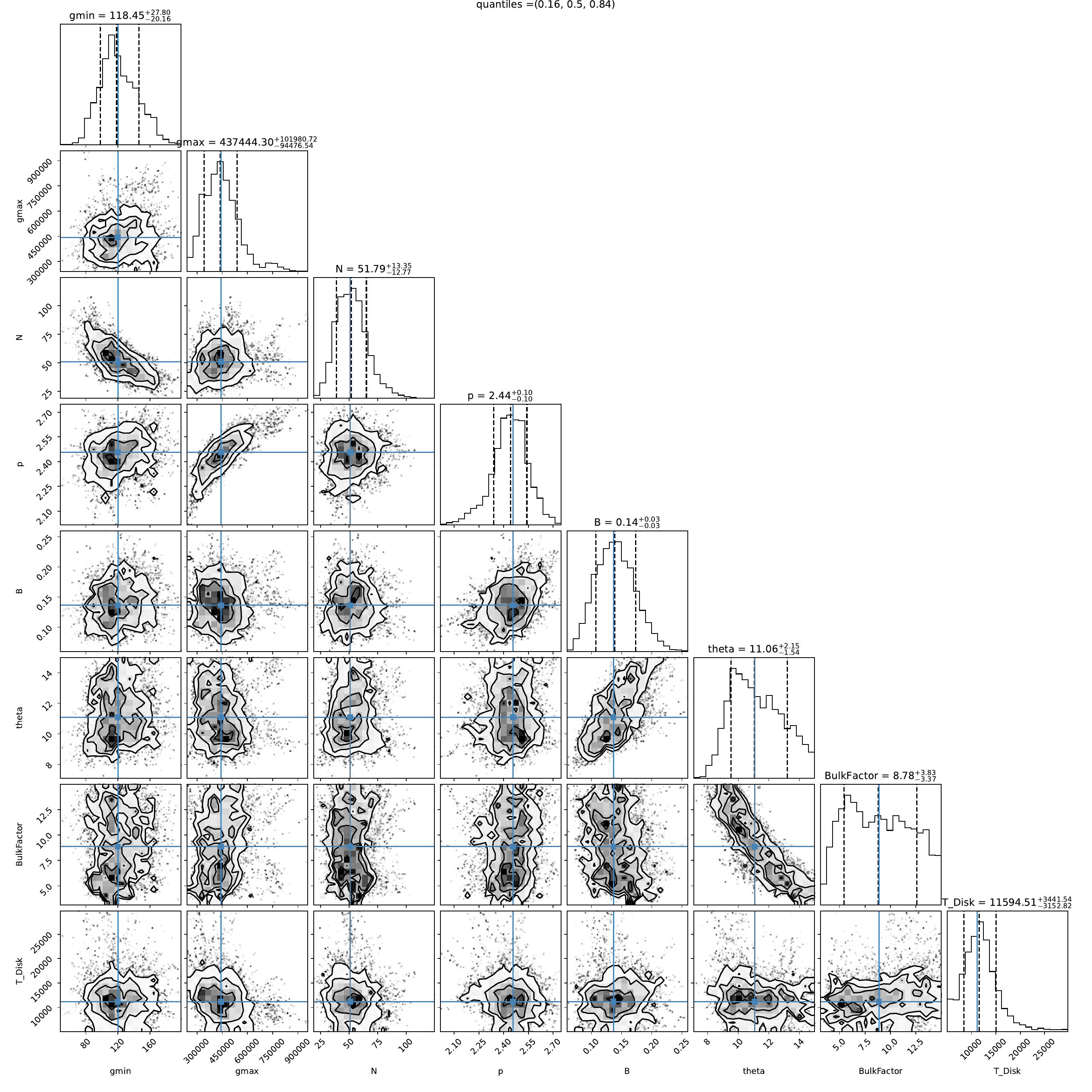}
\caption{Corner plot summary of the parameter distributions estimated through the MCMC procedure for the low state SED model. The panels on the diagonal show the distribution of each parameter, indicating the best fit value and the derived uncertainties with the black solid and dashed lines, respectively. The best fit value of each parameter and its corresponding uncertainties are also indicated on top of each histogram. The panels below the diagonal show the 2-dimensional projections of the histograms for each pair of parameters.} 

\label{fig:SED_model_low_cornerplot}
\end{figure*}

In order to evaluate the statistical uncertainties and confidence bands of the models, we have performed a Markov chain Monte Carlo (MCMC) approach implemented in \texttt{JetSeT} making use of the \texttt{emcee} package \citep{foreman2013}. The best fit obtained from the SED modelling for each parameter is used for the MCMC procedure as the prior value. Then the asymmetric errors and the $\sim$3$\sigma$ (5-95\%) confidence level contours of the fit of each parameter are calculated through the MCMC approach with 500 iterations using 50 burn in steps and a number of walkers equal to 10 times the number of free parameters in each case, for a total of 100 and 150 walkers for the high and low state SEDs, respectively. The corresponding high- and low-state models with the 3$\sigma$ significance contours as derived from the MCMC method are displayed in Fig.~\ref{fig:SED_models_mc}. Moreover, the uncertainties of the derived parameters from both models are reported in corner plots resulting from the MCMC simulations, shown in Figs.~\ref{fig:SED_model_high_cornerplot} and \ref{fig:SED_model_low_cornerplot}. 

We note that the uncertainties reported in Figs.~\ref{fig:SED_model_high_cornerplot} and \ref{fig:SED_model_low_cornerplot} are representative of the set of parameters used in these two specific models. In fact, as a caveat, we note that typically these SED models are based on a set of a larger number of parameters, typically rather unconstrained. This leads to a degeneracy on the solution, with possibly different combinations of parameters leading to a statistically adequate solution in terms of goodness of the fit. Therefore, these uncertainties are not representative of the errors of other parameter combinations. In addition, we note that not all statistically possible solutions may have a physically meaningful interpretation. Therefore, an evaluation of the best fit parameters after the modelling is needed to validate whether the combination of parameters is physically suitable for describing the fitted SEDs and the broadband emission of the source.

\hfill

\newpage

\hfill

\newpage

\hfill

\newpage

\section{Comparison with a leptonic SSC model}\label{sec:appendixD}
\begin{figure*}[h]
    \subfigure{\includegraphics[width=0.49\textwidth]{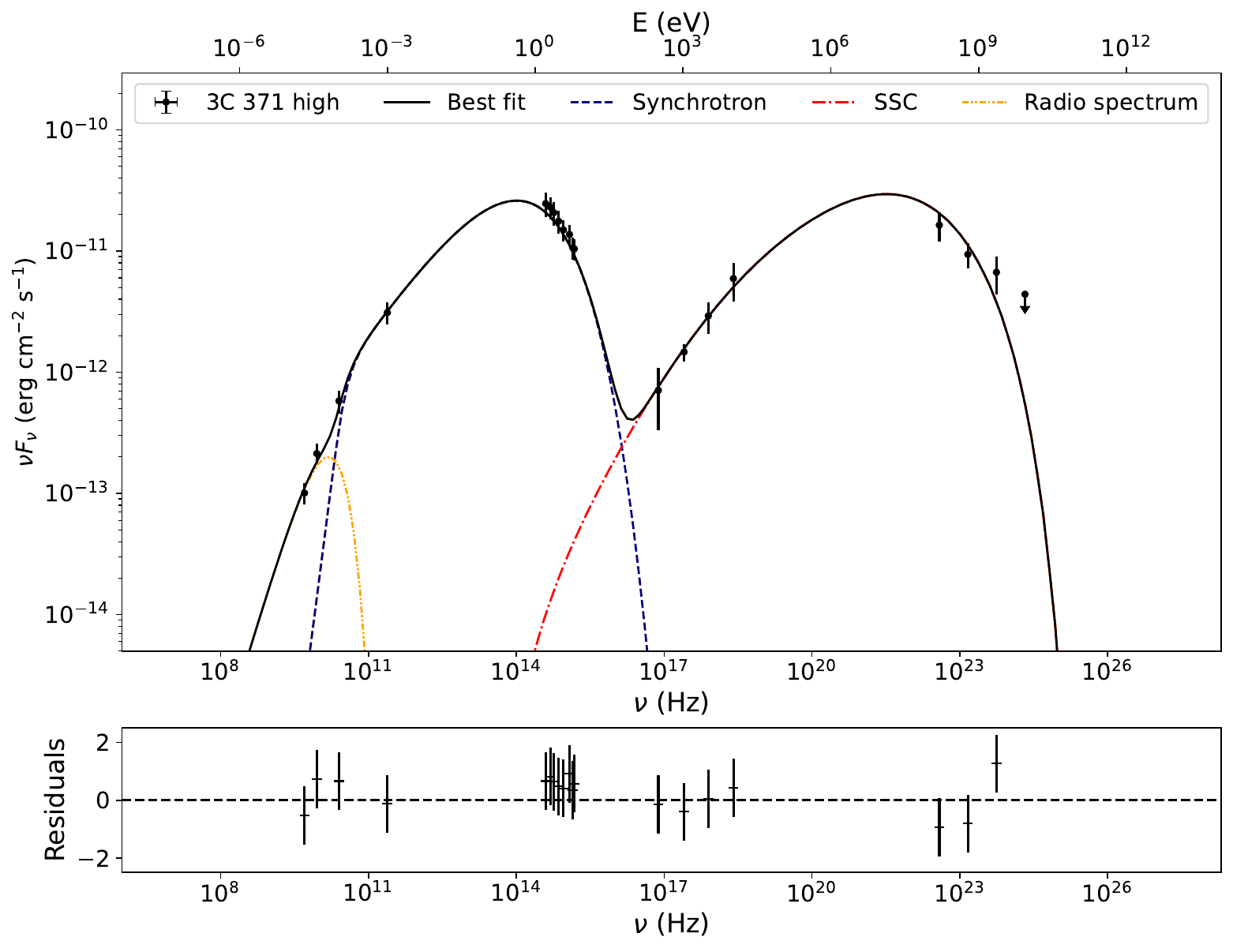}}
    \subfigure{\includegraphics[width=0.49\textwidth]{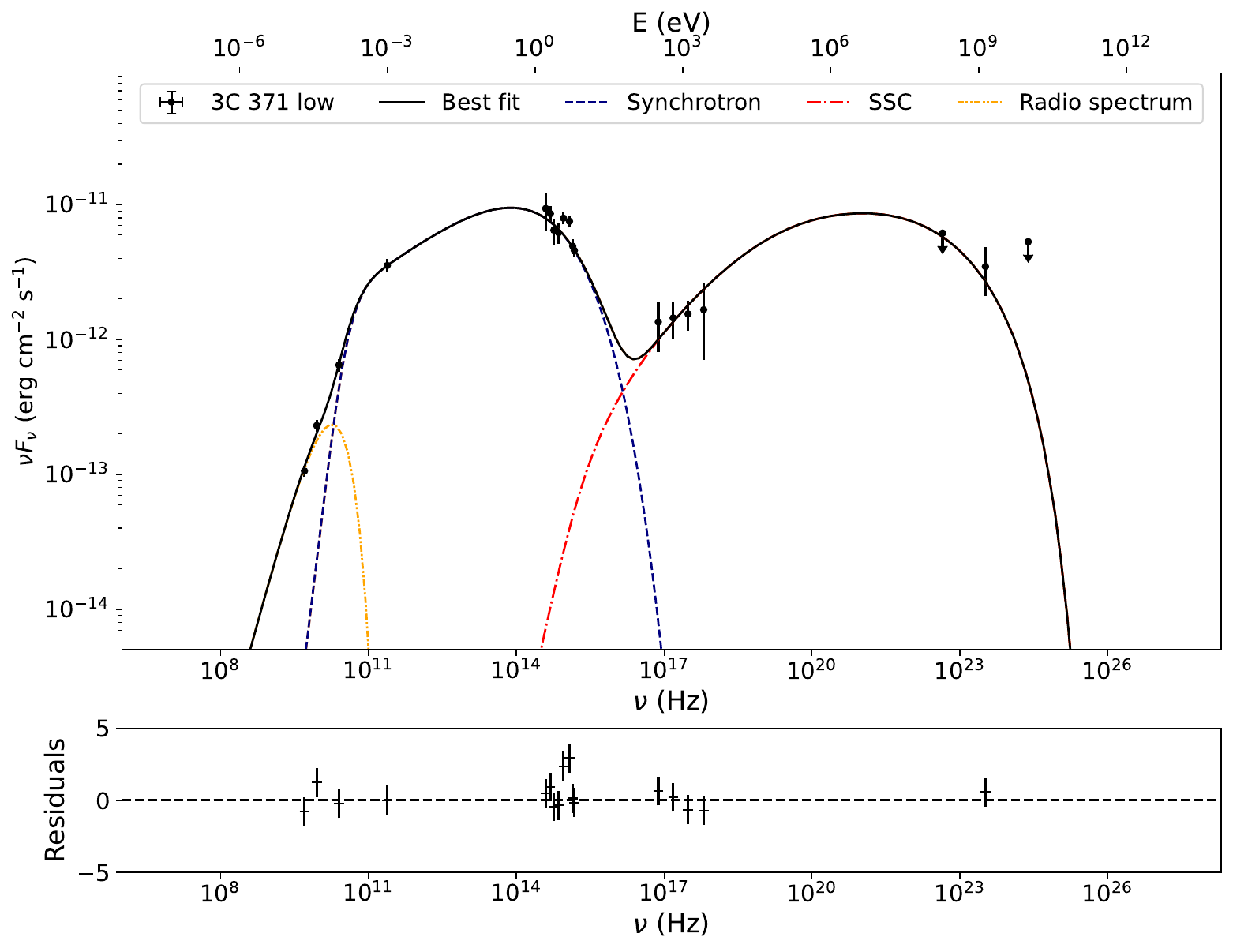}}
    \caption{Broadband SED models for the high (\textit{left}) and low (\textit{right}) brightness states of 3C~371 considering only $\gamma$-ray emission produced through SSC scattering. The top panels contain the broadband SEDs, with the contribution of each component and the best fit model being represented with different markers, lines and colours as indicated in the legend. The bottom panels represent the residuals of the fitted models.} 
    \label{fig:SED_models_SSC}
\end{figure*}

{In this section we include as a comparison the models obtained assuming a simpler leptonic SSC model, without including the contributions of the low-energy BLR and DT photons to the observed $\gamma$-ray emission via EC scattering. We also ignore here the thermal emission from the DT and AD contributing to the low-energy IR and optical emission, respectively. Therefore, we consider only the non-thermal emission of the jet. These models are shown in Fig.~\ref{fig:SED_models_SSC}.
}

{Comparing these models with those presented in Fig.~\ref{fig:SED_models} it is clear that the EC scattering contribution to the observed $\gamma$-ray emission is negligible in both emission states. Moreover, focusing on the IR-optical emission, we can also observe that during the high emission state, the DT and AD have also no impact on the total emission, being completely subdominant and opaqued by the much brighter synchrotron radiation from the relativistic jet. This is also true for the DT contribution during the low emission state.} 

{For the case of the AD in the latter state however, we see that there is a small difference with respect to the model including this component in the fit. The low-state model presented in Fig.~\ref{fig:SED_models} shows a slight excess in the UV bands associated with the AD. Despite still being subdominant and one order of magnitude below the synchrotron emission, according to our estimations based on the optical spectra of 3C 371 (see Sect.~\ref{sec7}), during the low state the AD contributes with roughly 10\% of the emission at UV frequencies, leading to this small excess. This becomes even clearer when comparing the residuals of both models. While for the model including the EC components, DT and AD, all residuals are well within an interval of $\pm$2, when considering only the SSC emission the UV residuals are clearly higher, especially in the UVOT $w1$ and $m2$ bands, where the emission from the AD peaks. This serves as additional confirmation of the presence of a faint AD in 3C 371, which becomes relevant when the synchrotron emission is faintest, as previously discussed.}

\end{appendix}

\end{document}